\documentclass[%
reprint,
amsmath,amssymb,
aip,
]{revtex4-2}

\usepackage{amsmath,amssymb,amsthm,color}
\newtheoremstyle{query}%
{}{}
{\color{red}}
{}
{\sffamily\bfseries}{:}{12pt}
{}
\theoremstyle{query}

\usepackage{graphicx}
\usepackage{dcolumn}
\usepackage{bm}

\begin{document}

\preprint{APS/123-QED}

\title{Studies of two-dimensional material resistive random-access memory by kinetic Monte Carlo simulations}

\author{Ying-Chuan Chen}
\affiliation{
Institute of Photonics and Optoelectronics, National Taiwan University, Taipei City 10617, Taiwan
}
\affiliation{
Corporate Research, Taiwan Semiconductor Manufacturing Co., Ltd., Hsinchu City 30078, Taiwan
}

\author{Yu-Ting Chao}
\affiliation{
Institute of Photonics and Optoelectronics, National Taiwan University, Taipei City 10617, Taiwan
}

\author{Edward Chen}
\affiliation{
Corporate Research, Taiwan Semiconductor Manufacturing Co., Ltd., Hsinchu City 30078, Taiwan
}

\author{Chao-Hsin Wu}
\altaffiliation{Author to whom correspondence should be addressed: chaohsinwu@ntu.edu.tw}
\affiliation{
Graduate Institute of Photonics and Optoelectronics, National Taiwan University, Taipei City 10617, Taiwan
}

\author{Yuh-Renn Wu}
\thanks{Author to whom correspondence should be addressed: yrwu@ntu.edu.tw}
\affiliation{
Graduate Institute of Photonics and Optoelectronics, National Taiwan University, Taipei City 10617, Taiwan
}

\date{\today}

\begin{abstract}
Resistive memory based on two-dimensional (2D) tungsten disulfide (WS$_2$), molybdenum disulfide (MoS$_2$), and hexagonal boron nitride (h-BN) materials is studied via experiments and simulations. The influence of the active layer thicknesses is discussed, and the thickness with the best On/Off ratio is found for 2D resistive random-access memory (RRAM). This work reveals fundamental differences between a 2D RRAM and conventional oxide RRAM. Furthermore, the physical parameters extracted using the kinetic Monte Carlo (KMC) model indicate that 2D materials have a lower diffusion activation energy along the vertical direction, where a smaller bias voltage and a shorter switching time can be achieved. The diffusion activation energy from the chemical vapor deposition (CVD)-grown sample is much lower than for mechanically exfoliated samples. The results suggest that MoS$_2$ has the fastest switching speed among the three considered 2D materials.
\end{abstract}

\maketitle


\section{Introduction}

The demand for storage devices is also growing along with the increasing popularity of artificial intelligence and the Internet of Things. However, traditional storage devices cannot meet the demand. For example, flash memory suffers from insufficient durability, the cache memory capacity is too small, etc. The most attention-grabbing group of new memory is called storage class memory, characterized by good access speed and improved cache memory capacity. RRAM is one example. Compared with traditional memory, RRAM has the advantages of high memory density ($\sim$2.5 times that of NOR flash memory (an electronic non-volatile memory made by the NOR logic gate)), high switching speed ($<10$ ns), and better durability ($>10^{6}$ times).\cite{yu2016emerging,li2014development}

Commonly used materials for RRAM are primarily transitional metal oxides, which are fabricated as a sandwich structure of metal/insulator/metal. This establishes the characteristic index of RRAM. The search for new RRAM materials has progressed in recent years to the utilization of 2D materials, such as graphene, h-BN, MoS$_2$, WS$_2$, and molybdenum ditelluride.\cite{singh2019enhanced,hou2019tubular,wang2018interface,li2019aerosol,zhang2018ultra}
Due to their low thermal budget, these 2D materials have the potential for back-end-of-line devices and monolithic three-dimensional integrated circuits.\cite{bablich2017graphene} The WS$_2$ and MoS$_2$ are currently the most promising 2D candidates for logic applications due to their high mobility and large bandgap. To effectively suppress the delay time and the power consumption between the logic and memory layers, the embedded memory made by WS$_2$/MoS$_2$ RRAM matches the requirements and can be further used for in-memory and neuromorphic computing.\cite{bishop2019monolithic} 

Here, WS$_2$ RRAM of different thicknesses is fabricated using gold/titanium (Au/Ti) contacts. The MoS$_2$ and h-BN 2D materials are also made for comparison. We applied the KMC method developed from Ginestra\cite{padovani2015microscopic} for experimental fitting to extract the material properties, which helps further determine the optimized structures in RRAM designs. Unlike analytic models, which often oversimplify the conduction mechanism, the physics-based KMC method coupled with Poisson and drift-diffusion solvers can extract the diffusion activation energy and obtain the defect formation along with their distributions. This provides an estimate of the thickness-dependent switching behaviors. The retention failure time and breakdown voltage predicted by the physics-based model agree well with the experiments. Furthermore, we find the thickness having the best On/Off ratio through the extracted WS$_2$ parameters and obtain the temperature-dependent characteristics. The ion transport properties for RRAM fabricated via CVD and mechanical exfoliation are compared. Finally, we consider the differences in the material properties of WS$_2$, MoS$_2$, and h-BN RRAM and discuss the feasibility of their application as 2D RRAM.

\begin{figure}
\centering
\includegraphics[width=\linewidth]{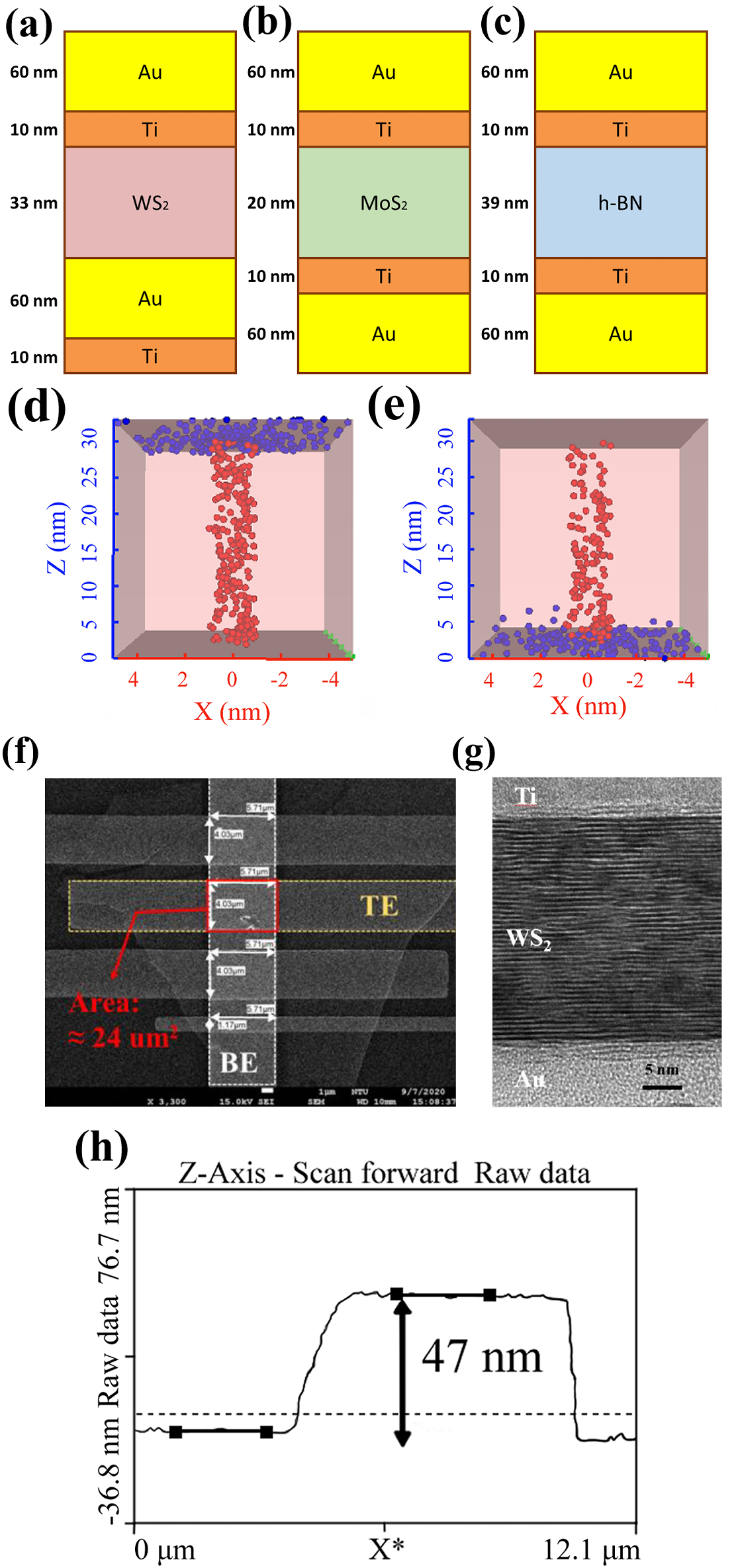}
\caption{Schematic diagrams for the (a) WS$_2$, (b) MoS$_2$, and (c) h-BN resistive random-access memory (RRAM) structures with plots for the 33 nm (d) WS$_2$ RRAM low resistance state (LRS) device structure and (e) WS$_2$ RRAM high resistance state (HRS) device structure. (f) The RRAM was fabricated by exfoliated WS$_2$, and the top electrode (TE)/bottom electrode (BE) overlap determined the conducting area, as shown in scanning electron microscopy analysis. (g) The WS$_2$ RRAM cross-section was analyzed via transmission electron microscopy. (h) The WS$_2$ thickness from the atomic force microscopy line scan.}
\label{fig:experiment}
\end{figure}

\section{Methods}

\subsection{Device fabrication and measurements}

This study performed experiments and simulations on WS$_{2}$, MoS$_{2}$, and h-BN RRAM. The device structures are shown in Figs.~\ref{fig:experiment}(a), \ref{fig:experiment}(b), and \ref{fig:experiment}(c), respectively, using Au/Ti for both the top electrode (TE) and bottom electrode (BE). To understand the impact of metal electrodes on device characteristics, the WS$_{2}$ was fabricated with asymmetric electrodes, and the MoS$_{2}$ and h-BN were fabricated with symmetric electrodes.

The active layer of the devices had various WS$_2$ thicknesses. Layers of 60 nm Au and 10 nm Ti were deposited on a heavily doped p-type silicon substrate as the BE of the RRAM device by electron-beam evaporation. The WS$_2$ flakes were obtained from single-crystal bulk material via mechanical exfoliation. The flakes were then transferred above the BE using polydimethylsiloxane. Afterward, the Au/Ti of 60/10 nm as the TE was deposited on the WS$_2$ flake to form an asymmetric WS$_2$ RRAM device. The cross-sectional area of the RRAM device was controlled by overlapping the TE and BE, as shown in Fig.~\ref{fig:experiment}(f). The vertical structure of the device was observed and measured via transmission electron microscopy and atomic force microscopy, as shown in Figs.~\ref{fig:experiment}(g) and \ref{fig:experiment}(h).

\subsection{Simulation methodology}

The switching state of the RRAM module was simulated to plot the I-V curves using the Ginestra\textsuperscript{TM} software. This software applies KMC modeling\cite{voter2007introduction,gillespie1976general} to simulate the generation, diffusion, and recombination of defects (vacancies and ions) in the RRAM device. The physics solver calculates the current value, including charge transport, temperature dependence, and 3D-space defects distribution. The KMC model simulates the dynamic defect distribution of the active layer. The physical models of the WS$_2$, MoS$_{2}$, and h-BN RRAM are based on experimental data to extract the physical parameters of the three 2D materials.

Figures \ref{fig:experiment}(d) and \ref{fig:experiment}(e) illustrate the device structures of the low resistance state (LRS) and high resistance state (HRS), respectively. The former has a complete conductive filament (CF), and the latter CF becomes sparser or even fractured because ions and vacancies recombine due to applied bias. Some defects that assist in carrier transport are recombined, causing the current to drop after reset operations. The current drop trend is highly correlated with the ion drift and diffusion in the lattice space. To shorten the RRAM simulation time, we only form one CF in the small cross-sectional area model, which causes the current drop to be discontinuous during reset operations. Therefore, the reset I-V curve in each device is the result of averaging many simulation curves.
There are multiple CFs formed on the large cross-sectional device, and the conduction states of each CF differ. Statistical averaging can be closer to the conduction state of the actual device.

The ion distribution in Figs.~\ref{fig:experiment}(d) or \ref{fig:experiment}(e) suggests that the ions gradually diffuse around CF, and the HRS resistance decreases progressively with subsequent resistance state switching. Therefore, ion diffusion in the in-plane direction is highly related to the device endurance. The 2D material has polarity, which is a typical bipolar switching mode in RRAM. Usually, the forming operation is applied with a positive bias, the set operation is in the forward bias, and the reset operation is in the reverse bias. If a negative bias is applied in the forming process, the set/reset switching voltages are reversed. Both positive and negative biases were applied in the forming operation of each device for the experiments, so the set/reset bias directions were opposite for various devices. A positive bias was uniformly applied in the simulations for the forming operations to facilitate subsequent analyses.

The vacancies and ions are generated in the device during forming or set operations by continuously increasing the applied bias. 
The generation rate is dependent on the 3D electric field in the device.\cite{mcpherson2003thermochemical,padovani2017microscopic} The associated Arrhenius equation is defined as
\begin{equation}
\\\\ R_{A,G}(X,Y,Z)=\nu \ exp[-\frac{E_{A,G}-p_{0}(2+\varepsilon_{r})/3\cdot F(X,Y,Z)}{k_{B}T}],
\label{eq_R_AG}
\end{equation}
where $\nu$ is a frequency prefactor, $E_{A,G}$ is the zero-field generation activation energy, $p_{0}$ is the polarizability, $\varepsilon_r$ is the relative permittivity, $k_B$ is Boltzmann's constant, and $T$ is the temperature. The electric field causes sulfur ions to drift in the active layer, while the diffusion rate depends on the local effective electric field along the diffusion direction. This Arrhenius equation is defined as
\begin{equation}
\\\\ R_{A,D}(X)=\nu \ exp[-\frac{E_{A,D}(X)-\gamma F_{X}}{k_{B}T}],
\label{eq_R_AD_X}
\end{equation}
\begin{equation}
\\\\ R_{A,D}(Y)=\nu \ exp[-\frac{E_{A,D}(Y)-\gamma F_{Y}}{k_{B}T}],
\label{eq_R_AD_Y}
\end{equation}
\begin{equation}
\\\\ R_{A,D}(Z)=\nu \ exp[-\frac{E_{A,D}(Z)-\gamma F_{Z}}{k_{B}T}],
\label{eq_R_AD_Z}
\end{equation}
where $E_{A,D}(X)$, $E_{A,D}(Y)$, and $E_{A,D}(Z)$ are respectively the diffusion activation energy in the X, Y, and Z direction, $\gamma$ is the field acceleration factor, and $F_{X}$, $F_{Y}$, and $F_{Z}$ are respectively the local electric field component of X, Y, and Z.

The retention time is how long a memory unit can retain a bit state at a specific temperature. For non-volatile memory, the most stringent requirement is retaining data for more than 10 years (about $\rm 3.1536 \times 10^{8}$ s) at operating temperatures up to 85 °C. To shorten the detection time, a device is placed in a high-temperature environment, and changes in time and resistance are recorded while baking. The Arrhenius diagram can be drawn by changing the temperature to record the retention failure time of the device and extract the activation energy. This is extrapolated to the working temperature to obtain the retention time at the considered temperature. An experimental report by Gao \textit{et al.}\cite{gao2010oxide} indicates that the retention time of the device is also calculated through the generation of the activation energy and theoretical formulas. The generation probability of the unbiased voltage is defined as
\begin{equation}
\\\\ p=exp(-E_{a}/k_{B}T),
\label{eq_p}
\end{equation}
where $E_a$ is the generation activation energy. The retention failure time of the device is defined as
\begin{equation}
\\\\ t_E=t_0/(n|ln(1-p)|)\approx t_0/np,
\label{eq_t_E}
\end{equation}
where $t_0$ is the oscillation period of lattice atoms, and $n$ is the number of escape directions for ions inside the lattice (for a cube, $n$ is 6). Usually, the generation probability will be far less than 1, so we apply a Taylor expansion to the original Eq.~(\ref{eq_t_E}) formula to obtain the first-order term. This prevents the dilemma where the denominator is zero and cannot be solved.

\section{Results and Discussion}

\subsection{Analysis of 2D RRAM characteristics}

\begin{figure}
\centering
\includegraphics[width=\linewidth]{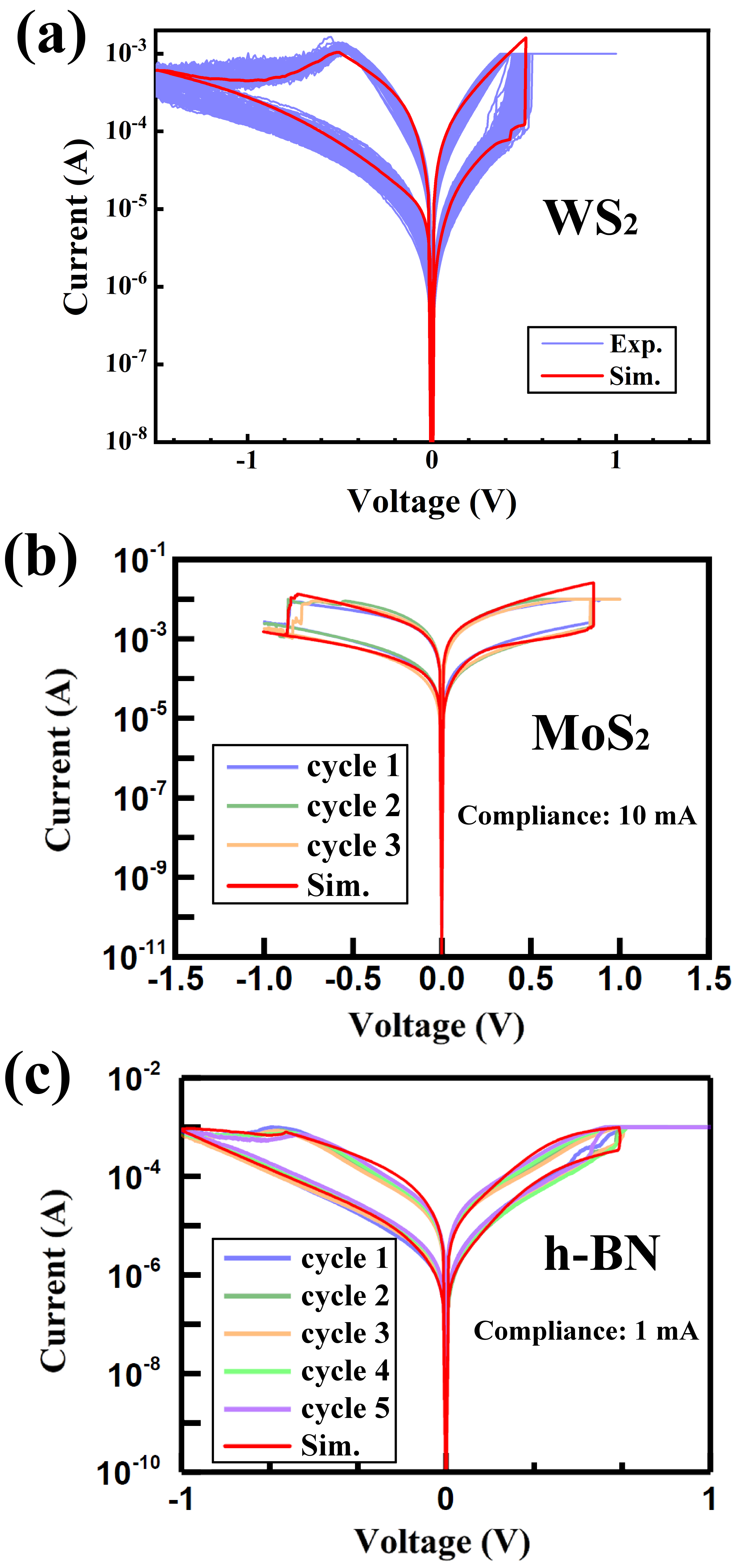}
\caption{(a) 33 nm WS$_2$, (b) 20 nm MoS$_2$,\cite{THWang2020} and (c) 39.4 nm h-BN RRAM set/reset I-V characteristics.\cite{THWang2020}}
\label{fig:WS2_RRAM}
\end{figure}

This work considered three 2D materials (WS$_2$, MoS$_2$, and h-BN) for comparison. More detailed studies on WS$_2$ materials have been made to build accurate models for KMC simulations, which can be used for device optimizations.
Figure \ref{fig:WS2_RRAM}(a) provides the measured and simulated set/reset current characteristics with 33-nm-thick WS$_2$ RRAM (44 layers), and the set/reset switching voltages are 0.5 and -0.6 V, respectively. The experimental data are the result of measuring 100 set/reset operation cycles, and the simulated set operation establishes a compliance current of $10^{-3}$ A. If the current is larger than $10^{-3}$ A during set switching, the system terminates the simulation and records the last data point. So, the simulated current after set operations is slightly larger than in the experimental data. 

The retention failure time is calculated by substituting the atomic oscillation period and the generation activation energy extracted from the simulations to Eq.~(\ref{eq_t_E}). The extracted generation activation energy of WS$_2$ is 1.11 eV, coinciding with the calculated formation energy of S vacancy by density functional theory (DFT).\cite{haldar2015systematic} The oscillation period is 18 fs,\cite{ho2017atomistic} and the retention failure time is $\rm 1.23\times10^4$ s at room temperature, or about 3 h. The experimental retention time of WS$_2$ can be maintained to $\rm 10^4$ s.
The experimental results of other researchers also show that the retention time of WS$_2$ can reach $\rm 10^4$ s.\cite{an2020highly, lee2019highly}

The simulation results indicate that the presence of an asymmetric electrode structure generates an internal electric field due to the utilization of different types of contact electrodes, thereby exerting an influence on the symmetry of the I-V curves. The work function of the contact surface with Au (BE) is larger than that of the Ti contact surface (TE) \cite{liu2022fermi}. Consequently, when the reverse bias voltage is swept, the internal electric field partially offsets the external electric field, leading to a lower current under the same bias voltage than the forward bias voltage.


Figure \ref{fig:WS2_RRAM}(b) gives the measured and simulated set/reset I-V characteristics of the 20-nm-thick MoS$_2$ RRAM. The switching voltages are 0.9 and -0.9 V, and the current of the reset switching dropping trend is faster than that of WS$ _2$. Thus, the field acceleration factor of MoS$_2$ is greater. The determined activation energy for the generation process of MoS2 stands at 1.13 eV, aligning harmoniously with the computationally determined formation energy of a sulfur (S) vacancy, as evaluated through DFT. \cite{haldar2015systematic} The oscillation period is 21.51 fs,\cite{ho2017effect} and the retention failure time at room temperature is estimated as $\rm 3.18\times10^4$ s, or about 8 h. The experimental results of other researchers also show that the retention time of MoS$_2$ can reach $\rm 10^4$ s.\cite{rehman2020decade}


Figure \ref{fig:WS2_RRAM}(c) provides the measured and simulated set/reset I-V characteristics of the 39.4-nm-thick h-BN RRAM. The switching voltages are 0.65 and -0.6 V, and the current of the reset switching dropping trend is smoother than that of WS$ _2$. Thus, the field acceleration factor of h-BN is lower. Taking into account the potential replacement of a nitrogen atom by oxygen in the surrounding environment, the extracted  generation activation energy of h-BN is 1.28 eV. This closes to the computationally derived formation energy of oxygen defects, as determined through the employment of the local density approximation (LDA).\cite{mosuang2002influence,PhysRevB.106.014107} The oscillation period is 24.4 fs,\cite{ghasemi2020mechanical} and the retention failure time at room temperature is estimated as $\rm 1.18\times10^7$ s, or about 136 days. Previous experimental results show that the retention time of h-BN reaches $\rm 10^7$ s.\cite{zhuang2020nonpolar} Although h-BN has a larger bandgap, the contact between h-BN and Ti causes the Fermi level of the electrode to be fixed on the p-type bandgap.\cite{bokdam2014schottky} The defects are generated in the depth of the bandgap due to the distribution of the Fermi level, and these vacancies can only transport a relatively small current. Therefore, the difference between the HRS and LRS current is small, and the On/Off ratio of h-BN cannot attain meaningful improvements.


\begin{table*}
\caption{\label{tab:basic_parameters}Basic parameters of the WS$_2$, MoS$_2$ and h-BN.}
\begin{ruledtabular}
\begin{tabular}{llccc}
Parameter & Description & WS$_2$ & MoS$_2$ & h-BN \\ 
\hline
$\epsilon_{r}$ & Relative permittivity & $6$\cite{roy2018electronic} & $7.1$\cite{davelou2014mos2} & $5.65$\cite{laturia2018dielectric} \\
$E_{g}$ (eV) & Bandgap & $1.54$\cite{roy2018electronic} & $1.23$\cite{gusakova2017electronic} & $5.97$\cite{bokdam2014schottky} \\ 
$E_{a}$ (eV) & Electron affinity & $3.92$\cite{roy2018electronic} & $4.2$\cite{xiao2018enhanced} & $0.8$\cite{bokdam2014schottky} \\
$k_{th}$ ($\rm W\cdot{cm}^{-1}\cdot K^{-1}$) & Thermal conductivity & $1.21$\cite{zhang2017systematic} & $0.035$\cite{gu2016layer} & $7.51$\cite{cai2019high} \\ 
$m_e$ ($\rm m_0$) & Electron density of states effective mass & $0.631$\cite{wickramaratne2014electronic} & $0.73$\cite{mishra2015screening} & $0.93$\cite{malozovsky2020accurate} \\
$m_h$ ($\rm m_0$) & Hole density of states effective mass & $0.832$\cite{wickramaratne2014electronic} & $0.78$\cite{mishra2015screening} & $0.77$\cite{malozovsky2020accurate} \\
\end{tabular}
\end{ruledtabular}
\end{table*}

\begin{table*}
\caption{\label{tab:MoS2_h-BN_parameters}Simulation parameters and switching times of the WS$_2$, MoS$_2$, h-BN, and HfO$\rm _x$\cite{loy2020oxygen} RRAM.}
\begin{ruledtabular}
\begin{tabular}{llcccc}
Parameter & Description & WS$_2$ & MoS$_2$ & h-BN & HfO$\rm _x$\cite{loy2020oxygen}\\
\hline
$E_{A,G}$ (eV) & Generation activation energy & $1.11$ & $1.13$ & $1.28$ & $2.9$ \\
$E_{A,D}$(X/Y) (eV) & Ion diffusion activation energy in the X/Y directions & $0.7$ & $0.7$ & $0.7$ & $0.7$ \\ 
$E_{A,D}$(Z) (eV) & Ion diffusion activation energy in the Z direction & $0.39$ & $0.2$ & $0.38$ & $0.7$ \\
$E_T$ (eV) & Thermal ionization energy of vacancy & $0.4\pm0.04$ & $0.3\pm0.04$ & $3.3\pm0.05$ & -- \\
$\nu$ (Hz) & Frequency prefactor & $4.5\times10^{13}$ & $4.5\times10^{13}$ & $4.5\times10^{13}$ & $7\times10^{13}$ \\
$p_0$ (e{\AA}) & Polarizability & $9$ & $29$ & $75$ & $5.2$ \\
$\gamma$ (e{\AA}) & Field acceleration factor & $0.2$ & $0.4$ & $0.01$ & $0.2$ \\
$t_{R}$ (sec) & Retention failure time & $1.23\times10^{4}$ & $3.18\times10^{4}$ & $1.18\times10^{7}$ & -- \\
$t_{S}$ (ns) & Switching time & $14.83$ & $7.33$ & $12.28$ & $21.33$ \\
\end{tabular}
\end{ruledtabular}
\end{table*}

Table \ref{tab:basic_parameters} gives the basic parameters used by three 2D materials in the simulations.
Table \ref{tab:MoS2_h-BN_parameters} indicates the simulation parameters used by the 2D materials and another hafnium oxide (HfO$\rm _x$) RRAM concept.\cite{loy2020oxygen} The WS$_2$ and MoS$_2$ have similar generation activation energies, and their retention failure times are of the same order. The diffusion activation energy of MoS$_2$ in the out-of-plane direction is lower than that of WS$_2$, and the field acceleration factor is higher. Thus, sulfur ions more easily drift and transport inside MoS$_2$, which is likely due to the mass densities of the material. The mass density of MoS$_2$ is 5.06 $\rm g/cm^3$,\cite{haynes2016crc} and WS$_2$ is 7.5 $\rm g/cm^ 3$.\cite{eagleson1994concise} This lessens the drift resistance of sulfur ions inside MoS$_2$ from atomic collisions, increasing the transport speed. Therein, the reset switching current drops more rapidly.

The generation activation energy of h-BN is the highest among the three 2D materials and has a long retention failure time.
Its field acceleration factor is much lower than WS$_2$ and MoS$_2$, causing the decreased reset switching to be relatively small. This parameter characteristic explains why h-BN produces threshold switching\cite{THWang2020} in the experiments. Therefore, h-BN must be applied with sufficient operating power to completely separate ions and vacancies, forming non-volatile memory with stable CFs. Otherwise, ions and vacancies can only be stretched outward to form electric dipoles. After removing the applied electric field, ions recombine with vacancies, exhibiting volatile memory characteristics.

\begin{figure}
\centering
\includegraphics[width=\linewidth]{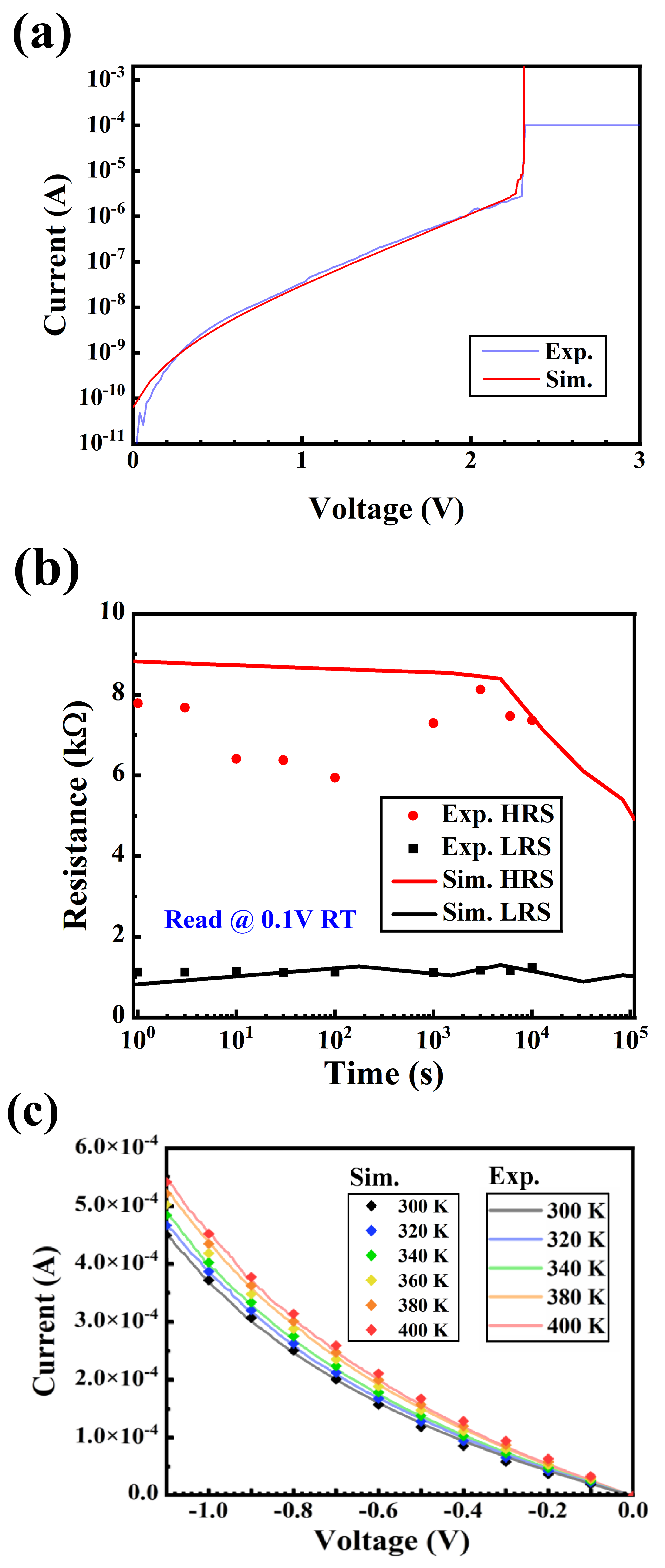}
\caption{(a) 12 nm WS$_2$ RRAM forming I-V characteristics,\cite{THWang2020} (b) WS$_2$ RRAM retention characteristics, and (c) WS$_2$ RRAM HRS temperature-dependent I-V characteristics.}
\label{fig:Forming}
\end{figure}

As HfO$\rm _x$ RRAM has a greater activation energy, the retention failure time at room temperature can exceed 10 years. Thus, HfO$\rm _x$ has been widely studied in RRAM applications. The On/Off ratio of WS$_2$ RRAM is about 10 at 0.1 V, while that of the HfO$\rm _x$ RRAM is about 50.\cite{loy2020oxygen} The On/Off ratio of WS$_2$ RRAM is five times smaller because its bandgap is smaller than HfO$\rm _x$. This increases the background current (not the current transported through the defect) of WS$_2$, resulting in a smaller difference between the HRS and LRS. Therein, the On/Off ratio of the overall device is relatively poor. Therefore, choosing a material with a large bandgap to make an RRAM can usually provide a higher On/Off ratio.

Compared with the benchmark device (HfO$\rm _x$ RRAM\cite{loy2020oxygen}), this work shows that 2D RRAM has a lower generation activation energy to generate defects at a smaller bias. The diffusion activation energy of 2D RRAM along the out-of-plane (Z) direction is also lower than that of HfO$\rm _x$ RRAM. This indicates that ions easily diffuse along the out-of-plane direction in 2D material compared to HfO$\rm _x$. No chemical bond between layers causes this effect. Under the 2D layered molecular arrangement, the torque perpendicular to the plane is more likely to cause molecular bond breaking and form defects, which act as channels for ion transport. Therefore, the set/reset switching voltage of 2D RRAM is lower than that of HfO$\rm _x$, which means 2D RRAM has a faster resistance switching speed. The data in Table \ref{tab:MoS2_h-BN_parameters} indicate that MoS$_2$ has the shortest switching time. Thus, MoS$_2$ is the most suitable for making RRAM devices among the three considered 2D materials.
To verify the reliability of the KMC model, we conduct a series of experiments and simulation comparisons for WS$_2$ in the next section.

\subsection{Comparison of WS$_2$ RRAM experiments and simulations}

\subsubsection{Forming operations}

Figure \ref{fig:Forming}(a) provides the measured and simulated forming current characteristics of the RRAM device made from 12-nm-thick WS$_2$ (16 layers) flakes. The device begins to break down at 2.3 V, and the defects inside the material generate larger quantities, causing the current to rise rapidly. In the experiments, a compliance current prevents the over-reaction of the material and causes the device to burn. The current is limited to $10^{-4}$ A when the applied voltage exceeds 2.3 V. The simulations calculate the current characteristics of the device region that has not been burned, so the current continues to increase. After the forming operation, the device model is output on the compliance current to continue the reset and set simulations.

\subsubsection{Retention time}

Figure \ref{fig:Forming}(b) gives the measured and simulated data at a reading voltage of 0.1 V. The solid dots are the HRS and LRS resistances measured in the experiments, which only recorded to $\rm 10^4$ s. The retention time of WS$_2$ was maintained at least to $\rm 10^ 4$ s. The solid line is the resistance calculated in the simulations. The actual device reduces the HRS resistance due to switching multiple times. Therefore, it is reasonable that the simulated On/Off ratio is slightly larger than the experimental result. In Fig.~\ref{fig:Forming}(b), the simulation results predict a significant drop in the HRS resistance after $\rm 10^4$ s. The resistance state cannot maintain stability. This agrees well with previous retention failure times ($\rm 1.23\times10^4$ s) calculated by Eq.~(\ref{eq_t_E}). It is speculated that the resistance of the HRS experimental data falling to 6 k$\Omega$ has a high probability due to experimental measurement error. The LRS experimental and simulation data show a stable resistance state at 1 k$\Omega$.

\subsubsection{Temperature-dependent analysis}

The electrical characteristics of the 24-nm-thick WS$_2$ (32 layers) were measured in the temperature-dependent experiments, where the state of this device was HRS. The specific heat density and thermal conductivity used to simulate the heat conduction are 1.19 $\rm J\cdot{cm}^{-3}\cdot K^{-1}$ and 1.21 $\rm W\cdot{cm}^ {-1}\cdot K^{-1}$.\cite{zhang2017systematic} The simulation and experimental data agree well, and the 360 K data missing from the experiment were added to Fig.~\ref{fig:Forming}(a) via simulation. The experimental trend of the adjacent temperature (340 and 380 K) indicates consistency in the corrected data. In the ambient temperature range from 300 to 400 K, the current increases proportionally with the temperature. When the applied bias increases, the current difference in the temperature in adjacent regions also increases. Thus, the temperature-dependent characteristics significantly affect the current. Therefore, we must consider changes in the ambient temperature on the current when WS$_2$ RRAM works under high bias.

\subsubsection{CVD-grown WS$_2$ RRAM analysis}

\begin{figure}
\centering
\includegraphics[width=\linewidth]{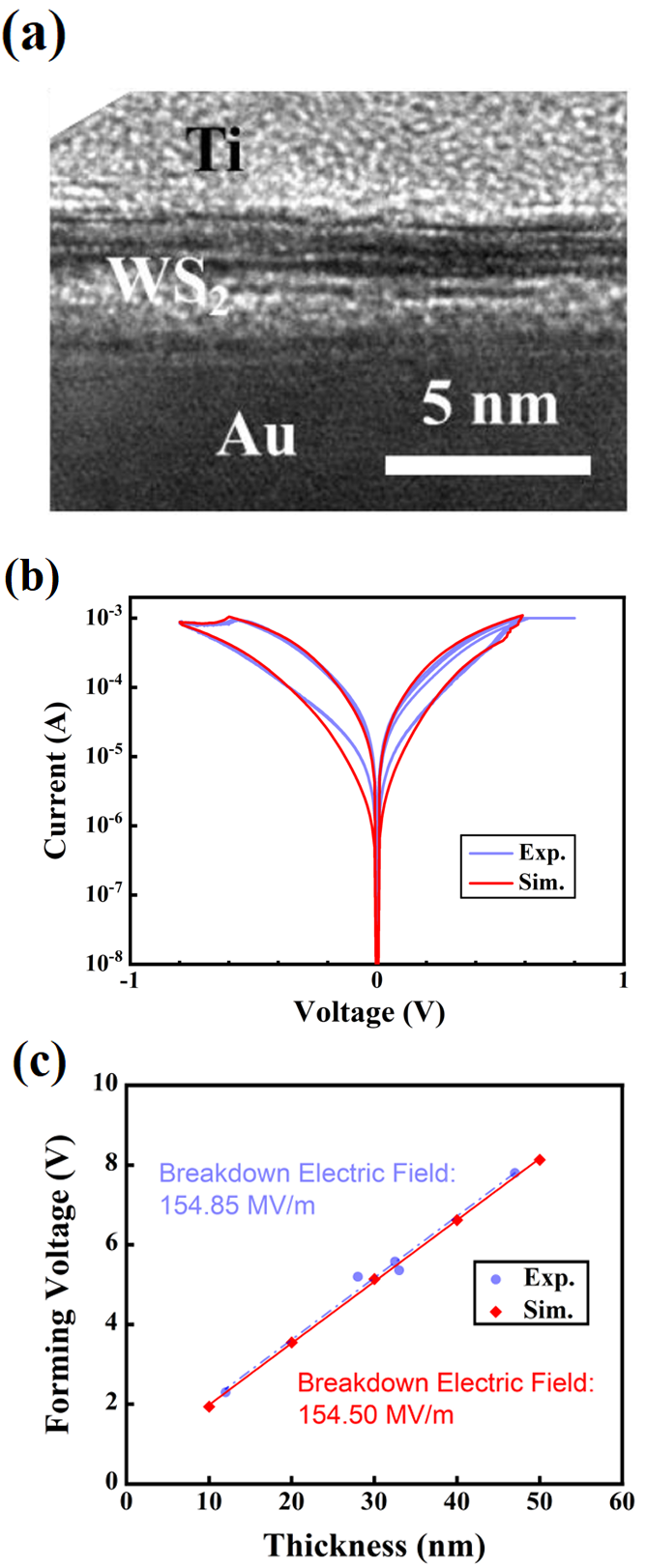}
\caption{(a) TEM was employed to analyze the CVD-grown large area transferred WS$_2$ RRAM. (b) CVD WS$_2$ RRAM set/reset I-V characteristics, and (c) correlation between the thickness and breakdown voltage of the WS$_2$ RRAM.}
\label{fig:CVD}
\end{figure}

This section considers differences in the electrical characteristics and material properties of WS$_2$ RRAM fabricated via mechanical exfoliation and CVD. The previous WS$_2$ RRAM was mechanically exfoliated to obtain the 2D flakes. Here, the device used CVD-grown WS$_2$ flakes, which are ultra-thin devices with a thickness of 3 nm (four layers), as shown in Fig.~\ref{fig:CVD}(a). Therefore, it is necessary to fine-tune some parameters in the simulations. We changed the basic parameters of the WS$_2$ bulk material to three or four layers. For example, the bandgap increased from the original 1.54 eV to 2.76 eV\cite{kim2015engineering} to simulate the carrier properties for the few-layer semiconductor device. The experimental and simulation results are shown in Fig.~\ref{fig:CVD}(b). The ultra-thin device strengthens the tunneling effect, so the overall current increases significantly with the applied bias.

\begin{table}
\caption{\label{tab:CVD_parameters}Simulation parameters and switching time of WS$_2$ RRAM with mechanical exfoliation and CVD.}
\begin{ruledtabular}
\begin{tabular}{lcc}
Parameter & Mechanical Exfoliation & CVD \\
\hline
$E_{A,G}$ (eV) & $1.11$ & $1.11$ \\
$E_{A,D}$(X/Y) (eV) & $0.7$ & $0.7$ \\
$E_{A,D}$(Z) (eV) & $0.39$ & $0.36$ \\
$E_T$ (eV) & $0.4\pm0.04$ & $0.75\pm0.04$ \\ 
$\nu$ (Hz) & $4.5\times10^{13}$ & $4.5\times10^{13}$ \\
$p_0$ (e{\AA}) & $9$ & $9$ \\
$\gamma$ (e{\AA}) & $0.2$ & $0.4$ \\
$t_{R}$ (sec) & $1.23\times10^{4}$ & $1.23\times10^{4}$ \\
$t_{S}$ (ns) & $14.83$ & $11.06$ \\
\end{tabular}
\end{ruledtabular}
\end{table}

Table \ref{tab:CVD_parameters} details the simulation parameters used in mechanical exfoliation and CVD WS$_2$ RRAM samples. The generation activation energy, polarizability, and frequency prefactor of the two samples are the same. This indicates that the generation of defects is independent of the arrangement state between layered molecules. There is a high correlation between the molecules' bond breaking and the induced electric field between polar molecules. The most apparent difference between the two fabrication methods is the drift and diffusion of sulfur ions inside the device. The WS$_2$ samples obtained via mechanical exfoliation are primarily single crystals, and the WS$_2$ grown via CVD is primarily polycrystalline. Therefore, CVD WS$_2$ has smaller domains and more defects. Each layer has more gaps and vacancies, and the probability of sulfur ion drifting and diffusion between the upper and lower layers is significantly increased.

Comparing the simulation results indicates that the CVD WS$_2$ RRAM has a lower out-of-plane diffusion activation energy and a higher field acceleration factor. Thus, it has a faster reset switching speed and a lower switching voltage. The CVD WS$_2$ device also has several defects. Therein, the required voltage for the forming operation is lower, and the switching time is shorter. In other words, the switching power consumption is lower. The experiments also confirm these results. Therefore, WS$_2$ obtained via CVD is more suitable for making RRAM devices than mechanical exfoliation.

\subsubsection{Breakdown electric field}

The experiment measured the breakdown voltages of five devices with thicknesses of 10, 20, 30, 40, and 50 nm for the RRAM devices from the simulations. We assume that the device is an ideal defect-free material with WS$_2$ parameters from Table \ref{tab:MoS2_h-BN_parameters}. The simulation results are consistent with the experimental data, as shown in Fig.~\ref{fig:CVD}(c).
The breakdown electric field is extracted through the WS$_2$ thickness and the breakdown voltage. The experimental and simulation results give a breakdown electric field for WS$_2$ of 155 MV/m. Defects are generated in the device when the internal electric field is greater than 155 MV/m, which causes the current to increase rapidly, and the initial forming operation is performed. We also used an initial model with few defects to simulate the forming operations of natural materials. The results indicate that the internal electric field generated by adding a few defects is insufficient to affect the breakdown voltage.

\subsection{Simulation predictions}

\begin{figure}
\centering
\includegraphics[width=\linewidth]{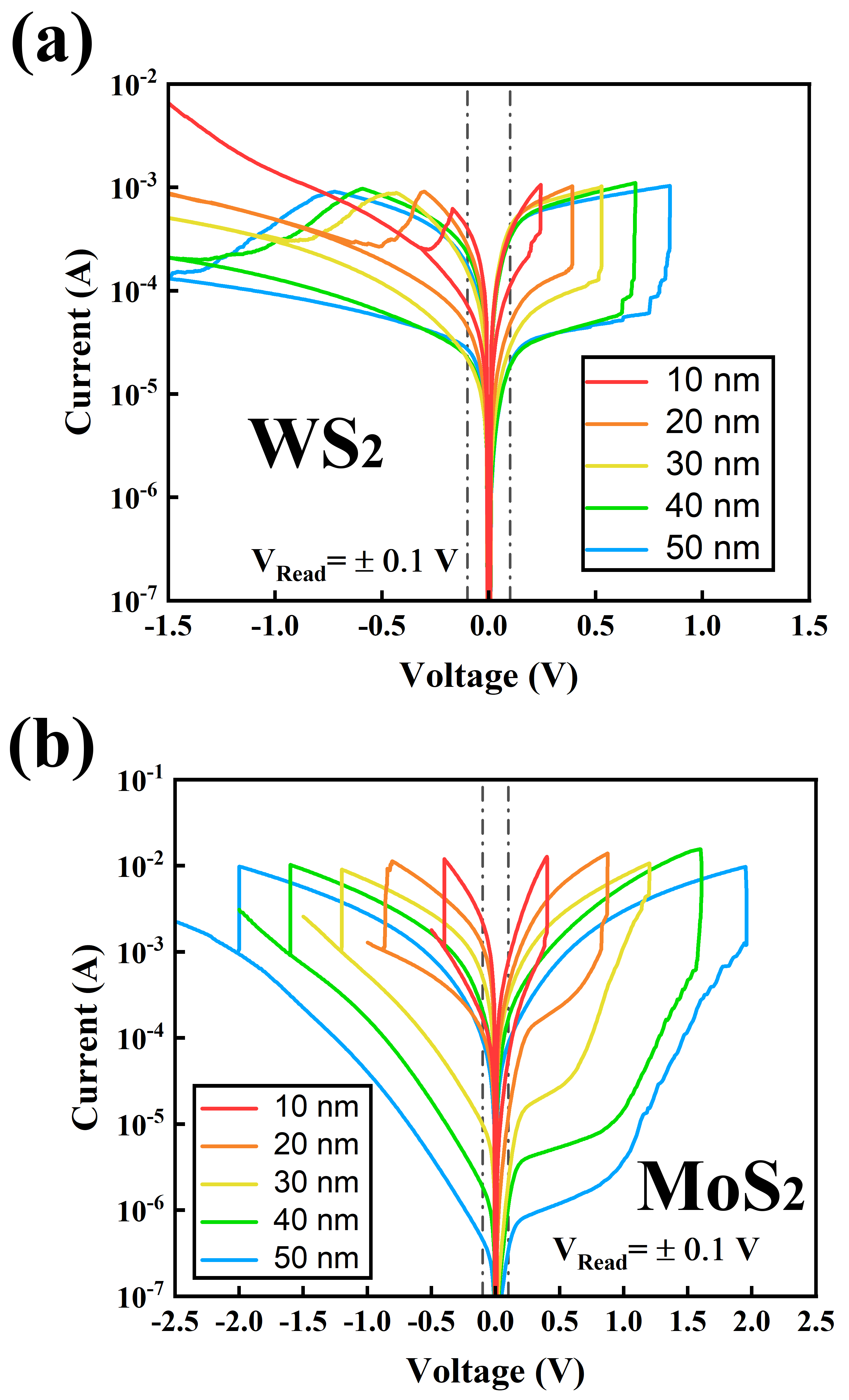}
\caption{Set/reset I-V characteristics of the (a) WS$_2$ and (b) MoS$_2$ RRAM devices with various thicknesses.}
\label{fig:thickness}
\end{figure}


The simulations are performed with the extracted WS$_2$ parameters to optimize the On/Off ratio as a function of thickness. The simulation results attain a stable I-V characteristic curve, as shown in Fig.~\ref{fig:thickness}(a). The stopping voltage and compliance current for each thickness are set to -1.5 V and $10^{-3}$ A. The TE and BE biases and the vertical electric field ($F=V/d$) indicate that thinner devices have smaller switching voltages, which is consistent with the simulation results. The current of thinner devices drops more rapidly during reset switching. The vertical electric field for thicker devices increases less from the sweeping bias, and the drift distance of ions in the vertical direction inside the device is longer. Thus, the overall ion drift speed is slower, and the switching time is longer, which gives a smoother current reduction curve. After all the ions drift from the TE to the BE, no more ions or vacancies exist for bonding recombination reactions. The current of the element then gradually increases with the bias voltage.

\begin{table}[ht]
\caption{\label{tab:on/off_ratio_0.1V_WS2}On/Off ratio for WS$_2$ RRAM at $\pm$0.1 V for various thicknesses.}
\begin{ruledtabular}
\begin{tabular}{ccc}
Thickness & On/Off Ratio & On/Off Ratio \\
(nm) & at -0.1 V & at 0.1 V \\
\hline
$10$ & $5.56$ & $3.20$ \\
$20$ & $6.14$ & $7.89$ \\ 
$30$ & $7.05$ & $14.43$ \\
$40$ & $10.13$ & $17.43$ \\ 
$50$ & $6.38$ & $17.10$ \\
\end{tabular}
\end{ruledtabular}
\end{table}


\begin{table}[ht]
\caption{\label{tab:on/off_ratio_0.1V_MoS2}On/Off ratio for MoS$_2$ RRAM at $\pm$0.1 V for various thicknesses.}
\begin{ruledtabular}
\begin{tabular}{cccc}
Thickness & On/Off Ratio & On/Off Ratio & Stopping Voltage\\
(nm) & at -0.1 V & at 0.1 V & (V)\\
\hline
$10$ & $12.98$ & $13.47$ & $-0.4$ \\
$20$ & $18.73$ & $32.66$ & $-0.86$ \\
$30$ & $51.39$ & $137.54$ & $-1.2$ \\
$40$ & $111.94$ & $149.16$ & $-1.6$ \\ 
$50$ & $208.68$ & $253.84$ & $-2.0$ \\
\end{tabular}
\end{ruledtabular}
\end{table}

The current trend for each thickness in Fig.~\ref{fig:thickness}(a) suggests that the current of the thicker device drops during reset operations because there are more defects. Both the differences in the currents and in the number of defects between the HRS and LRS increase. As the thickness exceeds 40 nm, ion diffusion in the in-plane direction and the vertical distance both increase, which reduces the difference in the number of defects. Thus, the HRS current tends to be at the same level as the LRS, and the On/Off ratio difference is small. Table \ref{tab:on/off_ratio_0.1V_WS2} provides the current and On/Off ratio of each thickness at -0.1 and 0.1 V biases. The 40-nm-thick device has the best On/Off ratio as it minimizes bit reading errors. Thinner devices have faster switching speeds and lower energy consumption, but their On/Off ratios are smaller, which explains why few-layer 2D materials are not commonly made into RRAM devices.

As MoS$_2$ has the fastest switching speed among the three considered 2D materials, we conduct electrical simulations for MoS$_2$ RRAM with various thicknesses. The simulation method is the same as in the previous section. The difference is the stopping voltage, which is set based on the state of the reset operations for each thickness. The stopping voltages for thicknesses of 10 to 50 nm are -0.5, -1.0, -1.5, -2.0, and -2.5 V so that the devices can be fully reset to attain the maximum on/off ratio.

Figure \ref{fig:thickness}(b) shows thicker devices have more extensive On/Off ratios. The On/Off ratios for each thickness are given in Table \ref{tab:on/off_ratio_0.1V_MoS2}. The 50 nm MoS$_2$ RRAM has the largest On/Off ratio, which is much larger than that of WS$_2$ RRAM. Therefore, MoS$_2$ is more suitable for RRAM devices in terms of switching speed and On/Off ratios.

The thickness of the RRAM should be determined based on the given operating voltage (V$_{DD}$) and power consumption constraints on the overall circuit system. Table \ref{tab:on/off_ratio_0.1V_MoS2} shows that thicker devices have greater switching voltages, which leads to increased power consumption. The stopping voltage of the 50 nm devices is set as a reverse bias voltage greater than 2.0 V. The current noise should be reduced to less than the device On/Off ratio to avoid memory misinterpretation.

\section{Summary}

We built physical models of WS$_2$, MoS$_2$, and h-BN RRAM through KMC simulations and experimental data to extract the physical parameters of these 2D materials. Theoretical formulas provided the retention failure times of WS$_2$, MoS$_2$, and h-BN RRAM as $\rm 1.23\times10^4$, $\rm 3.18\times10^4$, and $\rm 1.18\times10^7$ s, respectively. Compared with the HfO$\rm_x$ RRAM (benchmark device), the models show that ions easier drift in the out-of-plane direction due to the molecular arrangement of the 2D material. The 2D RRAM has a lower threshold voltage, increasing the switching speed. In particular, the switching speed of MoS$_2$ RRAM is the fastest of the three 2D RRAM devices. To verify the reliability of the KMC model, we comparatively analyzed the temperature-dependent experiments and simulations for WS$_2$ RRAMs. The current characteristics at high bias voltage are more significantly affected by temperature change. We discuss the physical attributes of RRAM made from 2D materials obtained via mechanical exfoliation and CVD. The results show that 2D materials grown via CVD have better device characteristics and are more suitable for making RRAM. Finally, electrical analysis and simulations for different active layer thicknesses provide the breakdown electric field of WS$_2$ as 155 MV/m. The thickness models simulate the On/Off ratio and switching voltage of WS$_2$ and MoS$_2$ RRAM at various thicknesses for reference in future 2D RRAM device designs.

If high-retention RRAM devices are desired, we suggest that the active layer be a material with high-generation activation energy (i.e., bond-dissociation energy) to improve retention. Although h-BN has a high retention failure time, its current difference is too low due to its low electron affinity. When h-BN is in contact with a metal, the Fermi level will be pinned to the p-type level. Thus, the h-BN RRAM can only generate deep-level defects to make a low On/Off ratio, indicating it is not an ideal RRAM material. The RRAM materials with better retention should be outside the scope of 2D materials due to the high bandgap, high electron affinity, and high bond-dissociation energy. To develop 2D RRAM devices, rapid ion drift in 2D materials helps manufacture RRAM devices with faster switching speeds. This advantage can solve the slow switching speed for traditional solid-state disks and the volatile memory property of DRAM. 2D RRAM can make up for the speed gap of the memory hierarchy.\cite{li2018phase} To find faster switching speed RRAMs, 2D RRAM should be considered with more in-depth experimental and simulation studies on MoS$_2$ in the future.

\section*{ACKNOWLEDGMENTS}
This work was supported by the National Science and Technology Council under Grant Nos. NSTC 111-2221-E-002-075, 111-2622-8-002-001, 112-2119-M-002-013,  and 111-2218-E-002-025. The authors would like to thanks Tzu-Heng Wang for his early work on preparing MoS$_2$ and h-BN samples.

\section*{AUTHOR DECLARATIONS}

\subsection*{Conflict of Interest}

The authors have no conflicts to disclose.

\subsection*{Author Contributions}

Yu-Ting Chao fabricated and measured the 2D devices. Ying-Chuan Chen simulated various devices and analyzed the data and wrote the manuscript. Chao-hsin Wu conduct the experiment work and Yuh-Renn Wu supervises the simulation. All the authors discussed the results and explanations.

\section*{DATA AVAILABILITY}

The data that support the findings of this study are available from the corresponding author upon reasonable request.


\section*{references}

\bibliography{samp}

\begin{thebibliography}{44}%
\makeatletter
\providecommand \@ifxundefined [1]{%
 \@ifx{#1\undefined}
}%
\providecommand \@ifnum [1]{%
 \ifnum #1\expandafter \@firstoftwo
 \else \expandafter \@secondoftwo
 \fi
}%
\providecommand \@ifx [1]{%
 \ifx #1\expandafter \@firstoftwo
 \else \expandafter \@secondoftwo
 \fi
}%
\providecommand \natexlab [1]{#1}%
\providecommand \enquote  [1]{``#1''}%
\providecommand \bibnamefont  [1]{#1}%
\providecommand \bibfnamefont [1]{#1}%
\providecommand \citenamefont [1]{#1}%
\providecommand \href@noop [0]{\@secondoftwo}%
\providecommand \href [0]{\begingroup \@sanitize@url \@href}%
\providecommand \@href[1]{\@@startlink{#1}\@@href}%
\providecommand \@@href[1]{\endgroup#1\@@endlink}%
\providecommand \@sanitize@url [0]{\catcode `\\12\catcode `\$12\catcode
  `\&12\catcode `\#12\catcode `\^12\catcode `\_12\catcode `\%12\relax}%
\providecommand \@@startlink[1]{}%
\providecommand \@@endlink[0]{}%
\providecommand \url  [0]{\begingroup\@sanitize@url \@url }%
\providecommand \@url [1]{\endgroup\@href {#1}{\urlprefix }}%
\providecommand \urlprefix  [0]{URL }%
\providecommand \Eprint [0]{\href }%
\providecommand \doibase [0]{https://doi.org/}%
\providecommand \selectlanguage [0]{\@gobble}%
\providecommand \bibinfo  [0]{\@secondoftwo}%
\providecommand \bibfield  [0]{\@secondoftwo}%
\providecommand \translation [1]{[#1]}%
\providecommand \BibitemOpen [0]{}%
\providecommand \bibitemStop [0]{}%
\providecommand \bibitemNoStop [0]{.\EOS\space}%
\providecommand \EOS [0]{\spacefactor3000\relax}%
\providecommand \BibitemShut  [1]{\csname bibitem#1\endcsname}%
\let\auto@bib@innerbib\@empty
\bibitem [{\citenamefont {Yu}\ and\ \citenamefont
  {Chen}(2016)}]{yu2016emerging}%
  \BibitemOpen
  \bibfield  {author} {\bibinfo {author} {\bibfnamefont {S.}~\bibnamefont
  {Yu}}\ and\ \bibinfo {author} {\bibfnamefont {P.-Y.}\ \bibnamefont {Chen}},\
  }\bibfield  {title} {\enquote {\bibinfo {title} {Emerging memory
  technologies: Recent trends and prospects},}\ }\href@noop {} {\bibfield
  {journal} {\bibinfo  {journal} {IEEE J. Solid-State Circuits}\ }\textbf
  {\bibinfo {volume} {8}},\ \bibinfo {pages} {43--56} (\bibinfo {year}
  {2016})}\BibitemShut {NoStop}%
\bibitem [{\citenamefont {Li}(2014)}]{li2014development}%
  \BibitemOpen
  \bibfield  {author} {\bibinfo {author} {\bibfnamefont {M.-T.}\ \bibnamefont
  {Li}},\ }\bibfield  {title} {\enquote {\bibinfo {title} {Development and
  challenges of the new non-volatile memory},}\ }\href@noop {} {\bibfield
  {journal} {\bibinfo  {journal} {Nano Commun. Netw.}\ }\textbf {\bibinfo
  {volume} {21}},\ \bibinfo {pages} {9--14} (\bibinfo {year}
  {2014})}\BibitemShut {NoStop}%
\bibitem [{\citenamefont {Singh}\ \emph {et~al.}(2019)\citenamefont {Singh},
  \citenamefont {Kumar}, \citenamefont {Kumar}, \citenamefont {Kumar},\ and\
  \citenamefont {Kumar}}]{singh2019enhanced}%
  \BibitemOpen
  \bibfield  {author} {\bibinfo {author} {\bibfnamefont {R.}~\bibnamefont
  {Singh}}, \bibinfo {author} {\bibfnamefont {R.}~\bibnamefont {Kumar}},
  \bibinfo {author} {\bibfnamefont {A.}~\bibnamefont {Kumar}}, \bibinfo
  {author} {\bibfnamefont {D.}~\bibnamefont {Kumar}},\ and\ \bibinfo {author}
  {\bibfnamefont {M.}~\bibnamefont {Kumar}},\ }\bibfield  {title} {\enquote
  {\bibinfo {title} {Enhanced resistive switching in graphene oxide based
  composite thin film for nonvolatile memory applications},}\ }\href@noop {}
  {\bibfield  {journal} {\bibinfo  {journal} {MRX}\ }\textbf {\bibinfo {volume}
  {6}},\ \bibinfo {pages} {105621} (\bibinfo {year} {2019})}\BibitemShut
  {NoStop}%
\bibitem [{\citenamefont {Hou}\ \emph {et~al.}(2019)\citenamefont {Hou},
  \citenamefont {Pan}, \citenamefont {Yu}, \citenamefont {Zhang}, \citenamefont
  {Huang}, \citenamefont {Mei}, \citenamefont {Zhang},\ and\ \citenamefont
  {Zhou}}]{hou2019tubular}%
  \BibitemOpen
  \bibfield  {author} {\bibinfo {author} {\bibfnamefont {X.}~\bibnamefont
  {Hou}}, \bibinfo {author} {\bibfnamefont {R.}~\bibnamefont {Pan}}, \bibinfo
  {author} {\bibfnamefont {Q.}~\bibnamefont {Yu}}, \bibinfo {author}
  {\bibfnamefont {K.}~\bibnamefont {Zhang}}, \bibinfo {author} {\bibfnamefont
  {G.}~\bibnamefont {Huang}}, \bibinfo {author} {\bibfnamefont
  {Y.}~\bibnamefont {Mei}}, \bibinfo {author} {\bibfnamefont {D.~W.}\
  \bibnamefont {Zhang}},\ and\ \bibinfo {author} {\bibfnamefont
  {P.}~\bibnamefont {Zhou}},\ }\bibfield  {title} {\enquote {\bibinfo {title}
  {Tubular 3d resistive random access memory based on rolled-up h-bn tube},}\
  }\href@noop {} {\bibfield  {journal} {\bibinfo  {journal} {Small}\ }\textbf
  {\bibinfo {volume} {15}},\ \bibinfo {pages} {1803876} (\bibinfo {year}
  {2019})}\BibitemShut {NoStop}%
\bibitem [{\citenamefont {Wang}\ \emph {et~al.}(2018)\citenamefont {Wang},
  \citenamefont {Tian}, \citenamefont {Zhao}, \citenamefont {Zhang},
  \citenamefont {Mao}, \citenamefont {Qiao}, \citenamefont {Pang},
  \citenamefont {Li}, \citenamefont {Yang},\ and\ \citenamefont
  {Ren}}]{wang2018interface}%
  \BibitemOpen
  \bibfield  {author} {\bibinfo {author} {\bibfnamefont {X.-F.}\ \bibnamefont
  {Wang}}, \bibinfo {author} {\bibfnamefont {H.}~\bibnamefont {Tian}}, \bibinfo
  {author} {\bibfnamefont {H.-M.}\ \bibnamefont {Zhao}}, \bibinfo {author}
  {\bibfnamefont {T.-Y.}\ \bibnamefont {Zhang}}, \bibinfo {author}
  {\bibfnamefont {W.-Q.}\ \bibnamefont {Mao}}, \bibinfo {author} {\bibfnamefont
  {Y.-C.}\ \bibnamefont {Qiao}}, \bibinfo {author} {\bibfnamefont
  {Y.}~\bibnamefont {Pang}}, \bibinfo {author} {\bibfnamefont {Y.-X.}\
  \bibnamefont {Li}}, \bibinfo {author} {\bibfnamefont {Y.}~\bibnamefont
  {Yang}},\ and\ \bibinfo {author} {\bibfnamefont {T.-L.}\ \bibnamefont
  {Ren}},\ }\bibfield  {title} {\enquote {\bibinfo {title} {Interface
  engineering with $\rm {MoS}_{2}-pd$ nanoparticles hybrid structure for a low
  voltage resistive switching memory},}\ }\href@noop {} {\bibfield  {journal}
  {\bibinfo  {journal} {Small}\ }\textbf {\bibinfo {volume} {14}},\ \bibinfo
  {pages} {1702525} (\bibinfo {year} {2018})}\BibitemShut {NoStop}%
\bibitem [{\citenamefont {Li}\ \emph {et~al.}(2019)\citenamefont {Li},
  \citenamefont {Sivan}, \citenamefont {Niu}, \citenamefont {Veluri},
  \citenamefont {Zamburg}, \citenamefont {Leong}, \citenamefont {Chand},
  \citenamefont {Samanta}, \citenamefont {Wang}, \citenamefont {Feng} \emph
  {et~al.}}]{li2019aerosol}%
  \BibitemOpen
  \bibfield  {author} {\bibinfo {author} {\bibfnamefont {Y.}~\bibnamefont
  {Li}}, \bibinfo {author} {\bibfnamefont {M.}~\bibnamefont {Sivan}}, \bibinfo
  {author} {\bibfnamefont {J.~X.}\ \bibnamefont {Niu}}, \bibinfo {author}
  {\bibfnamefont {H.}~\bibnamefont {Veluri}}, \bibinfo {author} {\bibfnamefont
  {E.}~\bibnamefont {Zamburg}}, \bibinfo {author} {\bibfnamefont
  {J.}~\bibnamefont {Leong}}, \bibinfo {author} {\bibfnamefont
  {U.}~\bibnamefont {Chand}}, \bibinfo {author} {\bibfnamefont
  {S.}~\bibnamefont {Samanta}}, \bibinfo {author} {\bibfnamefont
  {X.}~\bibnamefont {Wang}}, \bibinfo {author} {\bibfnamefont {X.}~\bibnamefont
  {Feng}}, \emph {et~al.},\ }\bibfield  {title} {\enquote {\bibinfo {title}
  {Aerosol jet printed $\rm {WSe}_{2}$ based rram on kapton suitable for
  flexible monolithic memory integration},}\ }in\ \href@noop {} {\emph
  {\bibinfo {booktitle} {2019 IEEE International Conference on Flexible and
  Printable Sensors and Systems (FLEPS)}}}\ (\bibinfo {organization} {IEEE},\
  \bibinfo {year} {2019})\ pp.\ \bibinfo {pages} {1--3}\BibitemShut {NoStop}%
\bibitem [{\citenamefont {Zhang}\ \emph {et~al.}(2018)\citenamefont {Zhang},
  \citenamefont {Zhang}, \citenamefont {Shrestha}, \citenamefont {Zhu},
  \citenamefont {Maize}, \citenamefont {Krylyuk}, \citenamefont {Shakouri},
  \citenamefont {Campbell}, \citenamefont {Cheung}, \citenamefont {Bendersky}
  \emph {et~al.}}]{zhang2018ultra}%
  \BibitemOpen
  \bibfield  {author} {\bibinfo {author} {\bibfnamefont {F.}~\bibnamefont
  {Zhang}}, \bibinfo {author} {\bibfnamefont {H.}~\bibnamefont {Zhang}},
  \bibinfo {author} {\bibfnamefont {P.}~\bibnamefont {Shrestha}}, \bibinfo
  {author} {\bibfnamefont {Y.}~\bibnamefont {Zhu}}, \bibinfo {author}
  {\bibfnamefont {K.}~\bibnamefont {Maize}}, \bibinfo {author} {\bibfnamefont
  {S.}~\bibnamefont {Krylyuk}}, \bibinfo {author} {\bibfnamefont
  {A.}~\bibnamefont {Shakouri}}, \bibinfo {author} {\bibfnamefont
  {J.}~\bibnamefont {Campbell}}, \bibinfo {author} {\bibfnamefont
  {K.}~\bibnamefont {Cheung}}, \bibinfo {author} {\bibfnamefont
  {L.}~\bibnamefont {Bendersky}}, \emph {et~al.},\ }\bibfield  {title}
  {\enquote {\bibinfo {title} {An ultra-fast multi-level $\rm {MoTe}_{2}$-based
  rram},}\ }in\ \href@noop {} {\emph {\bibinfo {booktitle} {2018 IEEE
  International Electron Devices Meeting (IEDM)}}}\ (\bibinfo {organization}
  {IEEE},\ \bibinfo {year} {2018})\ pp.\ \bibinfo {pages} {22--7}\BibitemShut
  {NoStop}%
\bibitem [{\citenamefont {Bablich}\ \emph {et~al.}(2017)\citenamefont
  {Bablich}, \citenamefont {Kataria}, \citenamefont {Passi},\ and\
  \citenamefont {Lemme}}]{bablich2017graphene}%
  \BibitemOpen
  \bibfield  {author} {\bibinfo {author} {\bibfnamefont {A.}~\bibnamefont
  {Bablich}}, \bibinfo {author} {\bibfnamefont {S.}~\bibnamefont {Kataria}},
  \bibinfo {author} {\bibfnamefont {V.}~\bibnamefont {Passi}},\ and\ \bibinfo
  {author} {\bibfnamefont {M.~C.}\ \bibnamefont {Lemme}},\ }\bibfield  {title}
  {\enquote {\bibinfo {title} {Graphene and two-dimensional materials:
  Extending silicon technology for the future?}}\ }in\ \href@noop {} {\emph
  {\bibinfo {booktitle} {Integrated Nanodevice and Nanosystem Fabrication}}}\
  (\bibinfo  {publisher} {Jenny Stanford Publishing},\ \bibinfo {year} {2017})\
  pp.\ \bibinfo {pages} {27--74}\BibitemShut {NoStop}%
\bibitem [{\citenamefont {Bishop}\ \emph {et~al.}(2019)\citenamefont {Bishop},
  \citenamefont {Wong}, \citenamefont {Mitra},\ and\ \citenamefont
  {Shulaker}}]{bishop2019monolithic}%
  \BibitemOpen
  \bibfield  {author} {\bibinfo {author} {\bibfnamefont {M.~D.}\ \bibnamefont
  {Bishop}}, \bibinfo {author} {\bibfnamefont {H.-S.~P.}\ \bibnamefont {Wong}},
  \bibinfo {author} {\bibfnamefont {S.}~\bibnamefont {Mitra}},\ and\ \bibinfo
  {author} {\bibfnamefont {M.~M.}\ \bibnamefont {Shulaker}},\ }\bibfield
  {title} {\enquote {\bibinfo {title} {Monolithic 3-d integration},}\
  }\href@noop {} {\bibfield  {journal} {\bibinfo  {journal} {IEEE Micro}\
  }\textbf {\bibinfo {volume} {39}},\ \bibinfo {pages} {16--27} (\bibinfo
  {year} {2019})}\BibitemShut {NoStop}%
\bibitem [{\citenamefont {Padovani}\ \emph {et~al.}(2015)\citenamefont
  {Padovani}, \citenamefont {Larcher}, \citenamefont {Pirrotta}, \citenamefont
  {Vandelli},\ and\ \citenamefont {Bersuker}}]{padovani2015microscopic}%
  \BibitemOpen
  \bibfield  {author} {\bibinfo {author} {\bibfnamefont {A.}~\bibnamefont
  {Padovani}}, \bibinfo {author} {\bibfnamefont {L.}~\bibnamefont {Larcher}},
  \bibinfo {author} {\bibfnamefont {O.}~\bibnamefont {Pirrotta}}, \bibinfo
  {author} {\bibfnamefont {L.}~\bibnamefont {Vandelli}},\ and\ \bibinfo
  {author} {\bibfnamefont {G.}~\bibnamefont {Bersuker}},\ }\bibfield  {title}
  {\enquote {\bibinfo {title} {Microscopic modeling of $\rm {HfO}_{x}$ rram
  operations: From forming to switching},}\ }\href@noop {} {\bibfield
  {journal} {\bibinfo  {journal} {IEEE T-ED}\ }\textbf {\bibinfo {volume}
  {62}},\ \bibinfo {pages} {1998--2006} (\bibinfo {year} {2015})}\BibitemShut
  {NoStop}%
\bibitem [{\citenamefont {Voter}(2007)}]{voter2007introduction}%
  \BibitemOpen
  \bibfield  {author} {\bibinfo {author} {\bibfnamefont {A.~F.}\ \bibnamefont
  {Voter}},\ }\bibfield  {title} {\enquote {\bibinfo {title} {Introduction to
  the kinetic monte carlo method},}\ }in\ \href@noop {} {\emph {\bibinfo
  {booktitle} {Radiation effects in solids}}}\ (\bibinfo  {publisher}
  {Springer},\ \bibinfo {year} {2007})\ pp.\ \bibinfo {pages}
  {1--23}\BibitemShut {NoStop}%
\bibitem [{\citenamefont {Gillespie}(1976)}]{gillespie1976general}%
  \BibitemOpen
  \bibfield  {author} {\bibinfo {author} {\bibfnamefont {D.~T.}\ \bibnamefont
  {Gillespie}},\ }\bibfield  {title} {\enquote {\bibinfo {title} {A general
  method for numerically simulating the stochastic time evolution of coupled
  chemical reactions},}\ }\href@noop {} {\bibfield  {journal} {\bibinfo
  {journal} {J. Comput. Phys.}\ }\textbf {\bibinfo {volume} {22}},\ \bibinfo
  {pages} {403--434} (\bibinfo {year} {1976})}\BibitemShut {NoStop}%
\bibitem [{\citenamefont {McPherson}\ \emph {et~al.}(2003)\citenamefont
  {McPherson}, \citenamefont {Kim}, \citenamefont {Shanware},\ and\
  \citenamefont {Mogul}}]{mcpherson2003thermochemical}%
  \BibitemOpen
  \bibfield  {author} {\bibinfo {author} {\bibfnamefont {J.}~\bibnamefont
  {McPherson}}, \bibinfo {author} {\bibfnamefont {J.}~\bibnamefont {Kim}},
  \bibinfo {author} {\bibfnamefont {A.}~\bibnamefont {Shanware}},\ and\
  \bibinfo {author} {\bibfnamefont {H.}~\bibnamefont {Mogul}},\ }\bibfield
  {title} {\enquote {\bibinfo {title} {Thermochemical description of dielectric
  breakdown in high dielectric constant materials},}\ }\href@noop {} {\bibfield
   {journal} {\bibinfo  {journal} {Appl. Phys. Lett.}\ }\textbf {\bibinfo
  {volume} {82}},\ \bibinfo {pages} {2121--2123} (\bibinfo {year}
  {2003})}\BibitemShut {NoStop}%
\bibitem [{\citenamefont {Padovani}\ \emph {et~al.}(2017)\citenamefont
  {Padovani}, \citenamefont {Gao}, \citenamefont {Shluger},\ and\ \citenamefont
  {Larcher}}]{padovani2017microscopic}%
  \BibitemOpen
  \bibfield  {author} {\bibinfo {author} {\bibfnamefont {A.}~\bibnamefont
  {Padovani}}, \bibinfo {author} {\bibfnamefont {D.}~\bibnamefont {Gao}},
  \bibinfo {author} {\bibfnamefont {A.}~\bibnamefont {Shluger}},\ and\ \bibinfo
  {author} {\bibfnamefont {L.}~\bibnamefont {Larcher}},\ }\bibfield  {title}
  {\enquote {\bibinfo {title} {A microscopic mechanism of dielectric breakdown
  in $\rm {SiO}_{2}$ films: An insight from multi-scale modeling},}\
  }\href@noop {} {\bibfield  {journal} {\bibinfo  {journal} {Journal of Applied
  physics}\ }\textbf {\bibinfo {volume} {121}},\ \bibinfo {pages} {155101}
  (\bibinfo {year} {2017})}\BibitemShut {NoStop}%
\bibitem [{\citenamefont {Gao}\ \emph {et~al.}(2010)\citenamefont {Gao},
  \citenamefont {Kang}, \citenamefont {Zhang}, \citenamefont {Sun},
  \citenamefont {Chen}, \citenamefont {Liu}, \citenamefont {Liu}, \citenamefont
  {Han}, \citenamefont {Wang}, \citenamefont {Fang} \emph
  {et~al.}}]{gao2010oxide}%
  \BibitemOpen
  \bibfield  {author} {\bibinfo {author} {\bibfnamefont {B.}~\bibnamefont
  {Gao}}, \bibinfo {author} {\bibfnamefont {J.}~\bibnamefont {Kang}}, \bibinfo
  {author} {\bibfnamefont {H.}~\bibnamefont {Zhang}}, \bibinfo {author}
  {\bibfnamefont {B.}~\bibnamefont {Sun}}, \bibinfo {author} {\bibfnamefont
  {B.}~\bibnamefont {Chen}}, \bibinfo {author} {\bibfnamefont {L.}~\bibnamefont
  {Liu}}, \bibinfo {author} {\bibfnamefont {X.}~\bibnamefont {Liu}}, \bibinfo
  {author} {\bibfnamefont {R.}~\bibnamefont {Han}}, \bibinfo {author}
  {\bibfnamefont {Y.}~\bibnamefont {Wang}}, \bibinfo {author} {\bibfnamefont
  {Z.}~\bibnamefont {Fang}}, \emph {et~al.},\ }\bibfield  {title} {\enquote
  {\bibinfo {title} {Oxide-based $\rm {RRAM}$: $\rm {Physical}$ based retention
  projection},}\ }in\ \href@noop {} {\emph {\bibinfo {booktitle} {2010
  Proceedings of the European Solid State Device Research Conference}}}\
  (\bibinfo {organization} {IEEE},\ \bibinfo {year} {2010})\ pp.\ \bibinfo
  {pages} {392--395}\BibitemShut {NoStop}%
\bibitem [{\citenamefont {Haldar}\ \emph {et~al.}(2015)\citenamefont {Haldar},
  \citenamefont {Vovusha}, \citenamefont {Yadav}, \citenamefont {Eriksson},\
  and\ \citenamefont {Sanyal}}]{haldar2015systematic}%
  \BibitemOpen
  \bibfield  {author} {\bibinfo {author} {\bibfnamefont {S.}~\bibnamefont
  {Haldar}}, \bibinfo {author} {\bibfnamefont {H.}~\bibnamefont {Vovusha}},
  \bibinfo {author} {\bibfnamefont {M.~K.}\ \bibnamefont {Yadav}}, \bibinfo
  {author} {\bibfnamefont {O.}~\bibnamefont {Eriksson}},\ and\ \bibinfo
  {author} {\bibfnamefont {B.}~\bibnamefont {Sanyal}},\ }\bibfield  {title}
  {\enquote {\bibinfo {title} {Systematic study of structural, electronic, and
  optical properties of atomic-scale defects in the two-dimensional transition
  metal dichalcogenides $\rm {MX}_{2}$ {(M= Mo, W; X= S, Se, Te)}},}\
  }\href@noop {} {\bibfield  {journal} {\bibinfo  {journal} {Phys. Rev. B}\
  }\textbf {\bibinfo {volume} {92}},\ \bibinfo {pages} {235408} (\bibinfo
  {year} {2015})}\BibitemShut {NoStop}%
\bibitem [{\citenamefont {Ho}\ \emph {et~al.}(2017{\natexlab{a}})\citenamefont
  {Ho}, \citenamefont {Bui}, \citenamefont {Phan}, \citenamefont {Kawazoe},\
  and\ \citenamefont {Le}}]{ho2017atomistic}%
  \BibitemOpen
  \bibfield  {author} {\bibinfo {author} {\bibfnamefont {T.~H.}\ \bibnamefont
  {Ho}}, \bibinfo {author} {\bibfnamefont {V.~Q.}\ \bibnamefont {Bui}},
  \bibinfo {author} {\bibfnamefont {T.~B.}\ \bibnamefont {Phan}}, \bibinfo
  {author} {\bibfnamefont {Y.}~\bibnamefont {Kawazoe}},\ and\ \bibinfo {author}
  {\bibfnamefont {H.~M.}\ \bibnamefont {Le}},\ }\bibfield  {title} {\enquote
  {\bibinfo {title} {Atomistic observation of the collision and migration of
  $\rm {Li}$ on $\rm {MoSe}_{2}$ and $\rm {WS}_{2}$ surfaces through ab initio
  molecular dynamics},}\ }\href@noop {} {\bibfield  {journal} {\bibinfo
  {journal} {PCCP}\ }\textbf {\bibinfo {volume} {19}},\ \bibinfo {pages}
  {27332--27342} (\bibinfo {year} {2017}{\natexlab{a}})}\BibitemShut {NoStop}%
\bibitem [{\citenamefont {An}\ \emph {et~al.}(2020)\citenamefont {An},
  \citenamefont {Lee}, \citenamefont {Lee}, \citenamefont {Wu}, \citenamefont
  {Koo},\ and\ \citenamefont {Kim}}]{an2020highly}%
  \BibitemOpen
  \bibfield  {author} {\bibinfo {author} {\bibfnamefont {H.}~\bibnamefont
  {An}}, \bibinfo {author} {\bibfnamefont {Y.~H.}\ \bibnamefont {Lee}},
  \bibinfo {author} {\bibfnamefont {J.~H.}\ \bibnamefont {Lee}}, \bibinfo
  {author} {\bibfnamefont {C.}~\bibnamefont {Wu}}, \bibinfo {author}
  {\bibfnamefont {B.~M.}\ \bibnamefont {Koo}},\ and\ \bibinfo {author}
  {\bibfnamefont {T.~W.}\ \bibnamefont {Kim}},\ }\bibfield  {title} {\enquote
  {\bibinfo {title} {Highly stable and flexible memristive devices based on
  polyvinylpyrrolidone: $\rm {WS}_{2}$ quantum dots},}\ }\href@noop {}
  {\bibfield  {journal} {\bibinfo  {journal} {Sci. Rep.}\ }\textbf {\bibinfo
  {volume} {10}},\ \bibinfo {pages} {1--8} (\bibinfo {year}
  {2020})}\BibitemShut {NoStop}%
\bibitem [{\citenamefont {Lee}\ \emph {et~al.}(2019)\citenamefont {Lee},
  \citenamefont {Wu}, \citenamefont {Sung}, \citenamefont {An},\ and\
  \citenamefont {Kim}}]{lee2019highly}%
  \BibitemOpen
  \bibfield  {author} {\bibinfo {author} {\bibfnamefont {J.~H.}\ \bibnamefont
  {Lee}}, \bibinfo {author} {\bibfnamefont {C.}~\bibnamefont {Wu}}, \bibinfo
  {author} {\bibfnamefont {S.}~\bibnamefont {Sung}}, \bibinfo {author}
  {\bibfnamefont {H.}~\bibnamefont {An}},\ and\ \bibinfo {author}
  {\bibfnamefont {T.~W.}\ \bibnamefont {Kim}},\ }\bibfield  {title} {\enquote
  {\bibinfo {title} {Highly flexible and stable resistive switching devices
  based on $\rm {WS}_{2}$ nanosheets: poly (methylmethacrylate)
  nanocomposites},}\ }\href@noop {} {\bibfield  {journal} {\bibinfo  {journal}
  {Sci. Rep.}\ }\textbf {\bibinfo {volume} {9}},\ \bibinfo {pages} {1--8}
  (\bibinfo {year} {2019})}\BibitemShut {NoStop}%
\bibitem [{\citenamefont {Wang}\ and\ \citenamefont {Wu}(2020)}]{THWang2020}%
  \BibitemOpen
  \bibfield  {author} {\bibinfo {author} {\bibfnamefont {T.-H.}\ \bibnamefont
  {Wang}}\ and\ \bibinfo {author} {\bibfnamefont {C.-H.}\ \bibnamefont {Wu}},\
  }\emph {\bibinfo {title} {The research of $\rm {2D}$ material-based Resistive
  Random Access Memory and $\rm {1T}$-$\rm {1R}$ cell}},\ \href@noop {}
  {Master's thesis},\ \bibinfo  {school} {National Taiwan University} (\bibinfo
  {year} {2020})\BibitemShut {NoStop}%
\bibitem [{\citenamefont {Liu}\ \emph {et~al.}(2022)\citenamefont {Liu},
  \citenamefont {Choi}, \citenamefont {Hwang}, \citenamefont {Yoo},\ and\
  \citenamefont {Sun}}]{liu2022fermi}%
  \BibitemOpen
  \bibfield  {author} {\bibinfo {author} {\bibfnamefont {X.}~\bibnamefont
  {Liu}}, \bibinfo {author} {\bibfnamefont {M.~S.}\ \bibnamefont {Choi}},
  \bibinfo {author} {\bibfnamefont {E.}~\bibnamefont {Hwang}}, \bibinfo
  {author} {\bibfnamefont {W.~J.}\ \bibnamefont {Yoo}},\ and\ \bibinfo {author}
  {\bibfnamefont {J.}~\bibnamefont {Sun}},\ }\bibfield  {title} {\enquote
  {\bibinfo {title} {Fermi level pinning dependent 2d semiconductor devices:
  challenges and prospects},}\ }\href@noop {} {\bibfield  {journal} {\bibinfo
  {journal} {Adv. Mater.}\ }\textbf {\bibinfo {volume} {34}},\ \bibinfo {pages}
  {2108425} (\bibinfo {year} {2022})}\BibitemShut {NoStop}%
\bibitem [{\citenamefont {Ho}\ \emph {et~al.}(2017{\natexlab{b}})\citenamefont
  {Ho}, \citenamefont {Dong}, \citenamefont {Kawazoe},\ and\ \citenamefont
  {Le}}]{ho2017effect}%
  \BibitemOpen
  \bibfield  {author} {\bibinfo {author} {\bibfnamefont {T.~H.}\ \bibnamefont
  {Ho}}, \bibinfo {author} {\bibfnamefont {H.~C.}\ \bibnamefont {Dong}},
  \bibinfo {author} {\bibfnamefont {Y.}~\bibnamefont {Kawazoe}},\ and\ \bibinfo
  {author} {\bibfnamefont {H.~M.}\ \bibnamefont {Le}},\ }\bibfield  {title}
  {\enquote {\bibinfo {title} {Effect of elasticity of the $\rm {MoS}_{2}$
  surface on $\rm {Li}$ atom bouncing and migration: Mechanism from $\rm {Ab}$
  initio molecular dynamic investigations},}\ }\href@noop {} {\bibfield
  {journal} {\bibinfo  {journal} {J. Phys. Chem. C}\ }\textbf {\bibinfo
  {volume} {121}},\ \bibinfo {pages} {1329--1338} (\bibinfo {year}
  {2017}{\natexlab{b}})}\BibitemShut {NoStop}%
\bibitem [{\citenamefont {Rehman}\ \emph {et~al.}(2020)\citenamefont {Rehman},
  \citenamefont {Rehman}, \citenamefont {Gul}, \citenamefont {Kim},
  \citenamefont {Karimov},\ and\ \citenamefont {Ahmed}}]{rehman2020decade}%
  \BibitemOpen
  \bibfield  {author} {\bibinfo {author} {\bibfnamefont {M.~M.}\ \bibnamefont
  {Rehman}}, \bibinfo {author} {\bibfnamefont {H.~M. M.~U.}\ \bibnamefont
  {Rehman}}, \bibinfo {author} {\bibfnamefont {J.~Z.}\ \bibnamefont {Gul}},
  \bibinfo {author} {\bibfnamefont {W.~Y.}\ \bibnamefont {Kim}}, \bibinfo
  {author} {\bibfnamefont {K.~S.}\ \bibnamefont {Karimov}},\ and\ \bibinfo
  {author} {\bibfnamefont {N.}~\bibnamefont {Ahmed}},\ }\bibfield  {title}
  {\enquote {\bibinfo {title} {Decade of $\rm {2D}$-materials-based $\rm
  {RRAM}$ devices: a review},}\ }\href@noop {} {\bibfield  {journal} {\bibinfo
  {journal} {STAM}\ }\textbf {\bibinfo {volume} {21}},\ \bibinfo {pages}
  {147--186} (\bibinfo {year} {2020})}\BibitemShut {NoStop}%
\bibitem [{\citenamefont {Mosuang}\ and\ \citenamefont
  {Lowther}(2002)}]{mosuang2002influence}%
  \BibitemOpen
  \bibfield  {author} {\bibinfo {author} {\bibfnamefont {T.}~\bibnamefont
  {Mosuang}}\ and\ \bibinfo {author} {\bibfnamefont {J.}~\bibnamefont
  {Lowther}},\ }\bibfield  {title} {\enquote {\bibinfo {title} {Influence of
  defects on the {h-BN to c-BN} transformation},}\ }\href@noop {} {\bibfield
  {journal} {\bibinfo  {journal} {Phys. Rev. B}\ }\textbf {\bibinfo {volume}
  {66}},\ \bibinfo {pages} {014112} (\bibinfo {year} {2002})}\BibitemShut
  {NoStop}%
\bibitem [{\citenamefont {Huang}\ \emph {et~al.}(2022)\citenamefont {Huang},
  \citenamefont {Grzeszczyk}, \citenamefont {Vaklinova}, \citenamefont
  {Watanabe}, \citenamefont {Taniguchi}, \citenamefont {Novoselov},\ and\
  \citenamefont {Koperski}}]{PhysRevB.106.014107}%
  \BibitemOpen
  \bibfield  {author} {\bibinfo {author} {\bibfnamefont {P.}~\bibnamefont
  {Huang}}, \bibinfo {author} {\bibfnamefont {M.}~\bibnamefont {Grzeszczyk}},
  \bibinfo {author} {\bibfnamefont {K.}~\bibnamefont {Vaklinova}}, \bibinfo
  {author} {\bibfnamefont {K.}~\bibnamefont {Watanabe}}, \bibinfo {author}
  {\bibfnamefont {T.}~\bibnamefont {Taniguchi}}, \bibinfo {author}
  {\bibfnamefont {K.~S.}\ \bibnamefont {Novoselov}},\ and\ \bibinfo {author}
  {\bibfnamefont {M.}~\bibnamefont {Koperski}},\ }\bibfield  {title} {\enquote
  {\bibinfo {title} {Carbon and vacancy centers in hexagonal boron nitride},}\
  }\href {https://doi.org/10.1103/PhysRevB.106.014107} {\bibfield  {journal}
  {\bibinfo  {journal} {Phys. Rev. B}\ }\textbf {\bibinfo {volume} {106}},\
  \bibinfo {pages} {014107} (\bibinfo {year} {2022})}\BibitemShut {NoStop}%
\bibitem [{\citenamefont {Ghasemi}, \citenamefont {Rutledge},\ and\
  \citenamefont {Yazdani}(2020)}]{ghasemi2020mechanical}%
  \BibitemOpen
  \bibfield  {author} {\bibinfo {author} {\bibfnamefont {H.}~\bibnamefont
  {Ghasemi}}, \bibinfo {author} {\bibfnamefont {J.~E.}\ \bibnamefont
  {Rutledge}},\ and\ \bibinfo {author} {\bibfnamefont {H.}~\bibnamefont
  {Yazdani}},\ }\bibfield  {title} {\enquote {\bibinfo {title} {Mechanical
  properties of defective cyanoethynyl ($\rm {2D}$ polyaniline--$\rm {C3N}$): A
  comparative molecular dynamics study versus graphene and hexagonal boron
  nitride},}\ }\href@noop {} {\bibfield  {journal} {\bibinfo  {journal}
  {Physica E Low Dimens. Syst. Nanostruct.}\ }\textbf {\bibinfo {volume}
  {121}},\ \bibinfo {pages} {114085} (\bibinfo {year} {2020})}\BibitemShut
  {NoStop}%
\bibitem [{\citenamefont {Zhuang}\ \emph {et~al.}(2020)\citenamefont {Zhuang},
  \citenamefont {Lin}, \citenamefont {Ahn}, \citenamefont {Catalano},
  \citenamefont {Chou}, \citenamefont {Roy}, \citenamefont {Quevedo-Lopez},
  \citenamefont {Colombo}, \citenamefont {Cai},\ and\ \citenamefont
  {Banerjee}}]{zhuang2020nonpolar}%
  \BibitemOpen
  \bibfield  {author} {\bibinfo {author} {\bibfnamefont {P.}~\bibnamefont
  {Zhuang}}, \bibinfo {author} {\bibfnamefont {W.}~\bibnamefont {Lin}},
  \bibinfo {author} {\bibfnamefont {J.}~\bibnamefont {Ahn}}, \bibinfo {author}
  {\bibfnamefont {M.}~\bibnamefont {Catalano}}, \bibinfo {author}
  {\bibfnamefont {H.}~\bibnamefont {Chou}}, \bibinfo {author} {\bibfnamefont
  {A.}~\bibnamefont {Roy}}, \bibinfo {author} {\bibfnamefont {M.}~\bibnamefont
  {Quevedo-Lopez}}, \bibinfo {author} {\bibfnamefont {L.}~\bibnamefont
  {Colombo}}, \bibinfo {author} {\bibfnamefont {W.}~\bibnamefont {Cai}},\ and\
  \bibinfo {author} {\bibfnamefont {S.~K.}\ \bibnamefont {Banerjee}},\
  }\bibfield  {title} {\enquote {\bibinfo {title} {Nonpolar resistive switching
  of multilayer-$\rm {hBN}$-based memories},}\ }\href@noop {} {\bibfield
  {journal} {\bibinfo  {journal} {Adv. Electron. Mater.}\ }\textbf {\bibinfo
  {volume} {6}},\ \bibinfo {pages} {1900979} (\bibinfo {year}
  {2020})}\BibitemShut {NoStop}%
\bibitem [{\citenamefont {Bokdam}\ \emph {et~al.}(2014)\citenamefont {Bokdam},
  \citenamefont {Brocks}, \citenamefont {Katsnelson},\ and\ \citenamefont
  {Kelly}}]{bokdam2014schottky}%
  \BibitemOpen
  \bibfield  {author} {\bibinfo {author} {\bibfnamefont {M.}~\bibnamefont
  {Bokdam}}, \bibinfo {author} {\bibfnamefont {G.}~\bibnamefont {Brocks}},
  \bibinfo {author} {\bibfnamefont {M.~I.}\ \bibnamefont {Katsnelson}},\ and\
  \bibinfo {author} {\bibfnamefont {P.~J.}\ \bibnamefont {Kelly}},\ }\bibfield
  {title} {\enquote {\bibinfo {title} {Schottky barriers at hexagonal boron
  nitride/metal interfaces: A first-principles study},}\ }\href@noop {}
  {\bibfield  {journal} {\bibinfo  {journal} {Phys. Rev. B}\ }\textbf {\bibinfo
  {volume} {90}},\ \bibinfo {pages} {085415} (\bibinfo {year}
  {2014})}\BibitemShut {NoStop}%
\bibitem [{\citenamefont {Roy}\ and\ \citenamefont
  {Bermel}(2018)}]{roy2018electronic}%
  \BibitemOpen
  \bibfield  {author} {\bibinfo {author} {\bibfnamefont {S.}~\bibnamefont
  {Roy}}\ and\ \bibinfo {author} {\bibfnamefont {P.}~\bibnamefont {Bermel}},\
  }\bibfield  {title} {\enquote {\bibinfo {title} {Electronic and optical
  properties of ultra-thin 2d tungsten disulfide for photovoltaic
  applications},}\ }\href@noop {} {\bibfield  {journal} {\bibinfo  {journal}
  {Sol. Energy Mater Sol. Cells}\ }\textbf {\bibinfo {volume} {174}},\ \bibinfo
  {pages} {370--379} (\bibinfo {year} {2018})}\BibitemShut {NoStop}%
\bibitem [{\citenamefont {Davelou}\ \emph {et~al.}(2014)\citenamefont
  {Davelou}, \citenamefont {Kopidakis}, \citenamefont {Kioseoglou},\ and\
  \citenamefont {Remediakis}}]{davelou2014mos2}%
  \BibitemOpen
  \bibfield  {author} {\bibinfo {author} {\bibfnamefont {D.}~\bibnamefont
  {Davelou}}, \bibinfo {author} {\bibfnamefont {G.}~\bibnamefont {Kopidakis}},
  \bibinfo {author} {\bibfnamefont {G.}~\bibnamefont {Kioseoglou}},\ and\
  \bibinfo {author} {\bibfnamefont {I.~N.}\ \bibnamefont {Remediakis}},\
  }\bibfield  {title} {\enquote {\bibinfo {title} {$\rm {MoS}_{2}$
  nanostructures: Semiconductors with metallic edges},}\ }\href@noop {}
  {\bibfield  {journal} {\bibinfo  {journal} {Solid State Commun.}\ }\textbf
  {\bibinfo {volume} {192}},\ \bibinfo {pages} {42--46} (\bibinfo {year}
  {2014})}\BibitemShut {NoStop}%
\bibitem [{\citenamefont {Laturia}, \citenamefont {Van~de Put},\ and\
  \citenamefont {Vandenberghe}(2018)}]{laturia2018dielectric}%
  \BibitemOpen
  \bibfield  {author} {\bibinfo {author} {\bibfnamefont {A.}~\bibnamefont
  {Laturia}}, \bibinfo {author} {\bibfnamefont {M.~L.}\ \bibnamefont {Van~de
  Put}},\ and\ \bibinfo {author} {\bibfnamefont {W.~G.}\ \bibnamefont
  {Vandenberghe}},\ }\bibfield  {title} {\enquote {\bibinfo {title} {Dielectric
  properties of hexagonal boron nitride and transition metal dichalcogenides:
  from monolayer to bulk},}\ }\href@noop {} {\bibfield  {journal} {\bibinfo
  {journal} {NPJ 2D Mater. Appl.}\ }\textbf {\bibinfo {volume} {2}},\ \bibinfo
  {pages} {1--7} (\bibinfo {year} {2018})}\BibitemShut {NoStop}%
\bibitem [{\citenamefont {Gusakova}\ \emph {et~al.}(2017)\citenamefont
  {Gusakova}, \citenamefont {Wang}, \citenamefont {Shiau}, \citenamefont
  {Krivosheeva}, \citenamefont {Shaposhnikov}, \citenamefont {Borisenko},
  \citenamefont {Gusakov},\ and\ \citenamefont {Tay}}]{gusakova2017electronic}%
  \BibitemOpen
  \bibfield  {author} {\bibinfo {author} {\bibfnamefont {J.}~\bibnamefont
  {Gusakova}}, \bibinfo {author} {\bibfnamefont {X.}~\bibnamefont {Wang}},
  \bibinfo {author} {\bibfnamefont {L.~L.}\ \bibnamefont {Shiau}}, \bibinfo
  {author} {\bibfnamefont {A.}~\bibnamefont {Krivosheeva}}, \bibinfo {author}
  {\bibfnamefont {V.}~\bibnamefont {Shaposhnikov}}, \bibinfo {author}
  {\bibfnamefont {V.}~\bibnamefont {Borisenko}}, \bibinfo {author}
  {\bibfnamefont {V.}~\bibnamefont {Gusakov}},\ and\ \bibinfo {author}
  {\bibfnamefont {B.~K.}\ \bibnamefont {Tay}},\ }\bibfield  {title} {\enquote
  {\bibinfo {title} {Electronic properties of bulk and monolayer $\rm {TMDs}$:
  theoretical study within $\rm {DFT}$ framework ($\rm {GVJ}$-2e method)},}\
  }\href@noop {} {\bibfield  {journal} {\bibinfo  {journal} {phys. stat. sol.}\
  }\textbf {\bibinfo {volume} {214}},\ \bibinfo {pages} {1700218} (\bibinfo
  {year} {2017})}\BibitemShut {NoStop}%
\bibitem [{\citenamefont {Xiao}\ \emph {et~al.}(2018)\citenamefont {Xiao},
  \citenamefont {Zhang}, \citenamefont {Chen}, \citenamefont {Xu},\ and\
  \citenamefont {Deng}}]{xiao2018enhanced}%
  \BibitemOpen
  \bibfield  {author} {\bibinfo {author} {\bibfnamefont {J.}~\bibnamefont
  {Xiao}}, \bibinfo {author} {\bibfnamefont {Y.}~\bibnamefont {Zhang}},
  \bibinfo {author} {\bibfnamefont {H.}~\bibnamefont {Chen}}, \bibinfo {author}
  {\bibfnamefont {N.}~\bibnamefont {Xu}},\ and\ \bibinfo {author}
  {\bibfnamefont {S.}~\bibnamefont {Deng}},\ }\bibfield  {title} {\enquote
  {\bibinfo {title} {Enhanced performance of a monolayer $\rm
  {MoS}_{2}/{WSe}_{2}$ heterojunction as a photoelectrochemical cathode},}\
  }\href@noop {} {\bibfield  {journal} {\bibinfo  {journal} {Nanomicro Lett.}\
  }\textbf {\bibinfo {volume} {10}},\ \bibinfo {pages} {1--9} (\bibinfo {year}
  {2018})}\BibitemShut {NoStop}%
\bibitem [{\citenamefont {Zhang}\ \emph {et~al.}(2017)\citenamefont {Zhang},
  \citenamefont {Xie}, \citenamefont {Ouyang},\ and\ \citenamefont
  {Chen}}]{zhang2017systematic}%
  \BibitemOpen
  \bibfield  {author} {\bibinfo {author} {\bibfnamefont {Z.}~\bibnamefont
  {Zhang}}, \bibinfo {author} {\bibfnamefont {Y.}~\bibnamefont {Xie}}, \bibinfo
  {author} {\bibfnamefont {Y.}~\bibnamefont {Ouyang}},\ and\ \bibinfo {author}
  {\bibfnamefont {Y.}~\bibnamefont {Chen}},\ }\bibfield  {title} {\enquote
  {\bibinfo {title} {A systematic investigation of thermal conductivities of
  transition metal dichalcogenides},}\ }\href@noop {} {\bibfield  {journal}
  {\bibinfo  {journal} {Int. j. heat mass transf.}\ }\textbf {\bibinfo {volume}
  {108}},\ \bibinfo {pages} {417--422} (\bibinfo {year} {2017})}\BibitemShut
  {NoStop}%
\bibitem [{\citenamefont {Gu}, \citenamefont {Li},\ and\ \citenamefont
  {Yang}(2016)}]{gu2016layer}%
  \BibitemOpen
  \bibfield  {author} {\bibinfo {author} {\bibfnamefont {X.}~\bibnamefont
  {Gu}}, \bibinfo {author} {\bibfnamefont {B.}~\bibnamefont {Li}},\ and\
  \bibinfo {author} {\bibfnamefont {R.}~\bibnamefont {Yang}},\ }\bibfield
  {title} {\enquote {\bibinfo {title} {Layer thickness-dependent phonon
  properties and thermal conductivity of $\rm {MoS}_{2}$},}\ }\href@noop {}
  {\bibfield  {journal} {\bibinfo  {journal} {J. Appl. Phys.}\ }\textbf
  {\bibinfo {volume} {119}},\ \bibinfo {pages} {085106} (\bibinfo {year}
  {2016})}\BibitemShut {NoStop}%
\bibitem [{\citenamefont {Cai}\ \emph {et~al.}(2019)\citenamefont {Cai},
  \citenamefont {Scullion}, \citenamefont {Gan}, \citenamefont {Falin},
  \citenamefont {Zhang}, \citenamefont {Watanabe}, \citenamefont {Taniguchi},
  \citenamefont {Chen}, \citenamefont {Santos},\ and\ \citenamefont
  {Li}}]{cai2019high}%
  \BibitemOpen
  \bibfield  {author} {\bibinfo {author} {\bibfnamefont {Q.}~\bibnamefont
  {Cai}}, \bibinfo {author} {\bibfnamefont {D.}~\bibnamefont {Scullion}},
  \bibinfo {author} {\bibfnamefont {W.}~\bibnamefont {Gan}}, \bibinfo {author}
  {\bibfnamefont {A.}~\bibnamefont {Falin}}, \bibinfo {author} {\bibfnamefont
  {S.}~\bibnamefont {Zhang}}, \bibinfo {author} {\bibfnamefont
  {K.}~\bibnamefont {Watanabe}}, \bibinfo {author} {\bibfnamefont
  {T.}~\bibnamefont {Taniguchi}}, \bibinfo {author} {\bibfnamefont
  {Y.}~\bibnamefont {Chen}}, \bibinfo {author} {\bibfnamefont {E.~J.}\
  \bibnamefont {Santos}},\ and\ \bibinfo {author} {\bibfnamefont {L.~H.}\
  \bibnamefont {Li}},\ }\bibfield  {title} {\enquote {\bibinfo {title} {High
  thermal conductivity of high-quality monolayer boron nitride and its thermal
  expansion},}\ }\href@noop {} {\bibfield  {journal} {\bibinfo  {journal} {Sci.
  Adv.}\ }\textbf {\bibinfo {volume} {5}},\ \bibinfo {pages} {eaav0129, 1--8}
  (\bibinfo {year} {2019})}\BibitemShut {NoStop}%
\bibitem [{\citenamefont {Wickramaratne}, \citenamefont {Zahid},\ and\
  \citenamefont {Lake}(2014)}]{wickramaratne2014electronic}%
  \BibitemOpen
  \bibfield  {author} {\bibinfo {author} {\bibfnamefont {D.}~\bibnamefont
  {Wickramaratne}}, \bibinfo {author} {\bibfnamefont {F.}~\bibnamefont
  {Zahid}},\ and\ \bibinfo {author} {\bibfnamefont {R.~K.}\ \bibnamefont
  {Lake}},\ }\bibfield  {title} {\enquote {\bibinfo {title} {Electronic and
  thermoelectric properties of few-layer transition metal dichalcogenides},}\
  }\href@noop {} {\bibfield  {journal} {\bibinfo  {journal} {Chem. Phys.}\
  }\textbf {\bibinfo {volume} {140}},\ \bibinfo {pages} {124710} (\bibinfo
  {year} {2014})}\BibitemShut {NoStop}%
\bibitem [{\citenamefont {Mishra}\ \emph {et~al.}(2015)\citenamefont {Mishra},
  \citenamefont {Smith}, \citenamefont {Liu}, \citenamefont {Zahid},
  \citenamefont {Zhu}, \citenamefont {Guo},\ and\ \citenamefont
  {Salahuddin}}]{mishra2015screening}%
  \BibitemOpen
  \bibfield  {author} {\bibinfo {author} {\bibfnamefont {V.}~\bibnamefont
  {Mishra}}, \bibinfo {author} {\bibfnamefont {S.}~\bibnamefont {Smith}},
  \bibinfo {author} {\bibfnamefont {L.}~\bibnamefont {Liu}}, \bibinfo {author}
  {\bibfnamefont {F.}~\bibnamefont {Zahid}}, \bibinfo {author} {\bibfnamefont
  {Y.}~\bibnamefont {Zhu}}, \bibinfo {author} {\bibfnamefont {H.}~\bibnamefont
  {Guo}},\ and\ \bibinfo {author} {\bibfnamefont {S.}~\bibnamefont
  {Salahuddin}},\ }\bibfield  {title} {\enquote {\bibinfo {title} {Screening in
  ultrashort (5 nm) channel $\rm {MoS}_{2}$ transistors: A full-band quantum
  transport study},}\ }\href@noop {} {\bibfield  {journal} {\bibinfo  {journal}
  {IEEE T-ED}\ }\textbf {\bibinfo {volume} {62}},\ \bibinfo {pages}
  {2457--2463} (\bibinfo {year} {2015})}\BibitemShut {NoStop}%
\bibitem [{\citenamefont {Malozovsky}\ \emph {et~al.}(2020)\citenamefont
  {Malozovsky}, \citenamefont {Bamba}, \citenamefont {Stewart}, \citenamefont
  {Franklin},\ and\ \citenamefont {Bagayoko}}]{malozovsky2020accurate}%
  \BibitemOpen
  \bibfield  {author} {\bibinfo {author} {\bibfnamefont {Y.}~\bibnamefont
  {Malozovsky}}, \bibinfo {author} {\bibfnamefont {C.}~\bibnamefont {Bamba}},
  \bibinfo {author} {\bibfnamefont {A.}~\bibnamefont {Stewart}}, \bibinfo
  {author} {\bibfnamefont {L.}~\bibnamefont {Franklin}},\ and\ \bibinfo
  {author} {\bibfnamefont {D.}~\bibnamefont {Bagayoko}},\ }\bibfield  {title}
  {\enquote {\bibinfo {title} {Accurate, ground state electronic and transport
  properties of h-bn},}\ }\href@noop {} {\bibfield  {journal} {\bibinfo
  {journal} {BAPS}\ }\textbf {\bibinfo {volume} {65}} (\bibinfo {year}
  {2020})}\BibitemShut {NoStop}%
\bibitem [{\citenamefont {Loy}\ \emph {et~al.}(2020)\citenamefont {Loy},
  \citenamefont {Dananjaya}, \citenamefont {Chakrabarti}, \citenamefont {Tan},
  \citenamefont {Chow}, \citenamefont {Toh},\ and\ \citenamefont
  {Lew}}]{loy2020oxygen}%
  \BibitemOpen
  \bibfield  {author} {\bibinfo {author} {\bibfnamefont {D.~J.~J.}\
  \bibnamefont {Loy}}, \bibinfo {author} {\bibfnamefont {P.~A.}\ \bibnamefont
  {Dananjaya}}, \bibinfo {author} {\bibfnamefont {S.}~\bibnamefont
  {Chakrabarti}}, \bibinfo {author} {\bibfnamefont {K.~H.}\ \bibnamefont
  {Tan}}, \bibinfo {author} {\bibfnamefont {S.~C.}\ \bibnamefont {Chow}},
  \bibinfo {author} {\bibfnamefont {E.~H.}\ \bibnamefont {Toh}},\ and\ \bibinfo
  {author} {\bibfnamefont {W.~S.}\ \bibnamefont {Lew}},\ }\bibfield  {title}
  {\enquote {\bibinfo {title} {Oxygen vacancy density dependence with a hopping
  conduction mechanism in multilevel switching behavior of $\rm
  {HfO}_{2}$-based resistive random access memory devices},}\ }\href@noop {}
  {\bibfield  {journal} {\bibinfo  {journal} {ACS Appl. Electron. Mater.}\
  }\textbf {\bibinfo {volume} {2}},\ \bibinfo {pages} {3160--3170} (\bibinfo
  {year} {2020})}\BibitemShut {NoStop}%
\bibitem [{\citenamefont {Haynes}, \citenamefont {Lide},\ and\ \citenamefont
  {Bruno}(2016)}]{haynes2016crc}%
  \BibitemOpen
  \bibfield  {author} {\bibinfo {author} {\bibfnamefont {W.~M.}\ \bibnamefont
  {Haynes}}, \bibinfo {author} {\bibfnamefont {D.~R.}\ \bibnamefont {Lide}},\
  and\ \bibinfo {author} {\bibfnamefont {T.~J.}\ \bibnamefont {Bruno}},\
  }\href@noop {} {\emph {\bibinfo {title} {$\rm {CRC}$ handbook of chemistry
  and physics}}}\ (\bibinfo  {publisher} {CRC press},\ \bibinfo {year}
  {2016})\BibitemShut {NoStop}%
\bibitem [{\citenamefont {Eagleson}\ \emph {et~al.}(1994)\citenamefont
  {Eagleson} \emph {et~al.}}]{eagleson1994concise}%
  \BibitemOpen
  \bibfield  {author} {\bibinfo {author} {\bibfnamefont {M.}~\bibnamefont
  {Eagleson}} \emph {et~al.},\ }\href@noop {} {\emph {\bibinfo {title} {Concise
  encyclopedia chemistry}}}\ (\bibinfo  {publisher} {Walter de Gruyter},\
  \bibinfo {year} {1994})\BibitemShut {NoStop}%
\bibitem [{\citenamefont {Kim}\ \emph {et~al.}(2015)\citenamefont {Kim},
  \citenamefont {Kim}, \citenamefont {Lee}, \citenamefont {Lee}, \citenamefont
  {Choi}, \citenamefont {Lee}, \citenamefont {Lee}, \citenamefont {Jhang},
  \citenamefont {Park}, \citenamefont {Cheong} \emph
  {et~al.}}]{kim2015engineering}%
  \BibitemOpen
  \bibfield  {author} {\bibinfo {author} {\bibfnamefont {H.-C.}\ \bibnamefont
  {Kim}}, \bibinfo {author} {\bibfnamefont {H.}~\bibnamefont {Kim}}, \bibinfo
  {author} {\bibfnamefont {J.-U.}\ \bibnamefont {Lee}}, \bibinfo {author}
  {\bibfnamefont {H.-B.}\ \bibnamefont {Lee}}, \bibinfo {author} {\bibfnamefont
  {D.-H.}\ \bibnamefont {Choi}}, \bibinfo {author} {\bibfnamefont {J.-H.}\
  \bibnamefont {Lee}}, \bibinfo {author} {\bibfnamefont {W.~H.}\ \bibnamefont
  {Lee}}, \bibinfo {author} {\bibfnamefont {S.~H.}\ \bibnamefont {Jhang}},
  \bibinfo {author} {\bibfnamefont {B.~H.}\ \bibnamefont {Park}}, \bibinfo
  {author} {\bibfnamefont {H.}~\bibnamefont {Cheong}}, \emph {et~al.},\
  }\bibfield  {title} {\enquote {\bibinfo {title} {Engineering optical and
  electronic properties of $\rm {WS}_{2}$ by varying the number of layers},}\
  }\href@noop {} {\bibfield  {journal} {\bibinfo  {journal} {ACS nano}\
  }\textbf {\bibinfo {volume} {9}},\ \bibinfo {pages} {6854--6860} (\bibinfo
  {year} {2015})}\BibitemShut {NoStop}%
\bibitem [{\citenamefont {Li}\ \emph {et~al.}(2018)\citenamefont {Li},
  \citenamefont {Chen}, \citenamefont {Wang},\ and\ \citenamefont
  {Sun}}]{li2018phase}%
  \BibitemOpen
  \bibfield  {author} {\bibinfo {author} {\bibfnamefont {X.-B.}\ \bibnamefont
  {Li}}, \bibinfo {author} {\bibfnamefont {N.-K.}\ \bibnamefont {Chen}},
  \bibinfo {author} {\bibfnamefont {X.-P.}\ \bibnamefont {Wang}},\ and\
  \bibinfo {author} {\bibfnamefont {H.-B.}\ \bibnamefont {Sun}},\ }\bibfield
  {title} {\enquote {\bibinfo {title} {Phase-change superlattice materials
  toward low power consumption and high density data storage: Microscopic
  picture, working principles, and optimization},}\ }\href@noop {} {\bibfield
  {journal} {\bibinfo  {journal} {Adv. Funct. Mater.}\ }\textbf {\bibinfo
  {volume} {28}},\ \bibinfo {pages} {1803380} (\bibinfo {year}
  {2018})}\BibitemShut {NoStop}%
\end{thebibliography}%

\end{document}